\documentclass[aps,pra,reprint,superscriptaddress,longbibliography]{revtex4-1}
\usepackage{braket}
\usepackage{amsmath}
\usepackage{graphicx}
\usepackage[
	colorlinks=true,
	linkcolor=blue,
	citecolor=blue,
	urlcolor=blue,
	breaklinks
]{hyperref}

\begin{document}

    \title{Dynamics of open quantum systems by interpolation of von Neumann and classical master equations, and its application to quantum annealing}

    \author{Tadashi Kadowaki}
    \email{tadakado@gmail.com}
    \noaffiliation
    \date{\today}
    \begin{abstract}
        We propose a method to interpolate dynamics of von Neumann and classical master equations with an arbitrary mixing parameter to investigate the thermal effects in quantum dynamics.
        The two dynamics are mixed by intervening to continuously modify their solutions, thus coupling them indirectly instead of directly introducing a coupling term.
        This maintains the quantum system in a pure state even after the introduction of thermal effects and obtains not only a density matrix but also a state vector representation.
        Further, we demonstrate that the dynamics of a two-level system can be rewritten as a set of standard differential equations, resulting in quantum dynamics that includes thermal relaxation.
        These equations are equivalent to the optical Bloch equations at the weak coupling and asymptotic limits, implying that the dynamics cause thermal effects naturally.
        Numerical simulations of ferromagnetic and frustrated systems support this idea.
        Finally, we use this method to study thermal effects in quantum annealing, revealing nontrivial performance improvements for a spin glass model over a certain range of annealing time.
        This result may enable us to optimize the annealing time of real annealing machines.
    \end{abstract}
    \maketitle

    \section{Introduction}
    Quantum annealing (QA) in the Ising model has been studied as a means for solving the quadratic unconstrained binary optimization (QUBO) problems using the power of quantum mechanics to explore all possible combinations of variables\cite{Kadowaki1998,Kadowaki1998a,Brooke1999,Farhi2001,Santoro2002,Santoro2006,Das2008,Morita2008,Tanaka2017}.
    The Ising model can be used as a general framework for representing the QUBO problems using spin-1/2 particles as binary variables and encoding arbitrary relations between them using $p$-body ($p \ge 2$) interactions and local fields\cite{Lucas2014}.

    A hardware implementation of QA by using superconducting flux qubits as spin-1/2 variables from D-Wave Systems has enabled us to solve the real-world problems\cite{Johnson2011}.
    However, thermal fluctuations reduce the coherence times of quantum systems and reduce our control over them, which invokes a query as to whether trajectories of the quantum dynamics are wiped off in the D-Wave machine due to decoherence.
    Detailed studies have observed that the system operated below 20 mK could distinguish between quantum and classical (thermo-)dynamics based on the quantum signatures of a specially designed Ising model\cite{Boixo2013,Albash2015}.
    It has also been depicted that the performance of open quantum systems can be improved by thermal relaxation after anticrossing to recover probabilities from the first excited state\cite{Amin2008,Venuti2017,Dickson2013}.
    Other roles of thermal fluctuations have also been studied, such as efficiently retrieving the ground state of an Ising system using noisy interactions at finite temperatures\cite{Nishimura2016}.
    Although thermal relaxation can provide a performance advantage in certain situations, lower operating temperatures are required for larger systems\cite{Albash2017}.

    For open quantum systems, both macroscopic (e.g., Bloch and optical Bloch equations\cite{Bloch1946,Arecchi1965}) and microscopic (e.g., Redfield and Lindblad equations\cite{Redfield1965,Kossakowski1972}) methods have been studied.
    Although our approach also exhibits other applications, we investigate the behavior of a quantum annealer in this study.
    We intend to drive quantum systems using von Neumann and classical master equations with an arbitrary mixing ratio (coupling constant) in order to incorporate both the quantum and thermal effects.
    These systems obey quantum dynamics at one extreme mixing ratio and classical dynamics at the other, with a mixture of the two appearing at some point between the ends.
    The mixed dynamics depicts two noteworthy features: it has a temperature parameter (from the classical master equation), and it is designed to keep the systems in pure states, unlike the dynamics of other open quantum systems, which are generally described by mixed states.

    We further obtain an analytical representation of the dynamics for a two-level system, and the results are observed to be equivalent to the optical Bloch equations\cite{Metcalf1999,Arecchi1965} at the weak coupling and asymptotic limits, implying that the method is valid and that it provides an interpretation of the optical Bloch equations.
    The ground-state probability is continuously differentiable as a function of the mixing ratio at every point and is continuous but not differentiable as a function of the temperature at one particular point.
    We assume that this non-differentiable point may also be present in many-body systems and may be associated with a phase transition in the parameter space; however, we cannot completely answer that question in this study.

    We conducted numerical simulations of multi-spin systems, which confirmed that the relaxation of a ferromagnetic system is similar to that of the two-level system.
    Additionally, we use the frustrated and highly-degenerate model proposed by Boixo \textit{et al.}\cite{Boixo2013} to observe smooth transitions of the ground-state probabilities among the degenerate states that depend on the relative strength of the quantum and classical effects in the dynamics and the relation between the probability distribution and mixing parameter.

    Finally, we investigate the Sherrington--Kirkpatrick (SK) model\cite{Sherrington1975}, which is an infinite-range spin glass model consisting of random interactions, to study the combinatorial optimization problems with many local minima.
    We observe that the performance of the classical dynamics is better than that of the quantum one for short annealing times, whereas the quantum dynamics performs better for long annealing times.
    In the first situation, the quantum system is too far from being adiabatic to follow its instantaneous ground state, while thermal relaxation is still effective in the classical system even in the case of short annealing times.
    We further confirm that QA can efficiently follow its instantaneous ground state and retrieve the optimal solution with high probability using an appropriate annealing time and schedule.

    A nontrivial result arising from this analysis is that the mixed dynamics depicts a better performance than either the purely quantum or classical dynamics for some intermediate annealing times.
    This occurs when the annealing time is close to the point where the quantum and classical performance curves intersect each other.
    Thermally-assisted performance improvements have already been reported by theoretical studies of two-level systems and dissipative quasi-free fermions\cite{Amin2008,Venuti2017} and experimental studies of multi-spin systems\cite{Dickson2013}.
    We demonstrate that this phenomenon occurs for specific parameter regions of the annealing time and quantum-classical mixing parameter.
    This result may enable us to identify the optimal annealing times for real annealers, although the mixing ratio (coupling constant) is not easy to control.

    This study is organized as follows.
    Section~\ref{sec_formulation} formulates our proposed dynamics based on quantum and classical dynamics.
    Section \ref{sec_tls} further deals with a particular case, namely the dynamics of a two-level system.
    Section~\ref{sec_simulation} conducts numerical simulations for the Hushimi--Temperley (ferromagnetic), quantum signature Hamiltonian, and Sherrington--Kirkpatrick models.
    Finally, Sec.~\ref{sec_summary} summarizes and discusses our results.

    \section{Formulation}
    \label{sec_formulation}
    Initially, we discuss a method to interpolate the dynamics of (quantum) von Neumann and classical master equations.
    To perform this, we introduce a mixing parameter $\alpha$ to control the ratios of the two dynamics: when $\alpha = 0$, the system obeys purely quantum dynamics; when $\alpha = 1$, it undergoes purely thermodynamic relaxation to reach an equilibrium state.
    We use a density matrix representation for the quantum dynamics.
    However, a state vector representation can also be used.
    Without loss of generality, we focus on the spin systems in this study, and the Hamiltonian ${\mathcal H}$ consists of a diagonal part ${\mathcal H}_c$ with $\sigma^z$ and an off-diagonal part ${\mathcal H}_q$ with $\sigma^x$ and/or $\sigma^y$.
    Only the diagonal part ${\mathcal H}_c$ is used in the classical master equation.
    Given the Hamiltonian of the quantum system, the time evolution of the density matrix $\rho$ can be described using the von Neumann equation,
    \begin{align}
        \frac{d \rho}{dt} & = - i [ {\mathcal H}, \rho ] \\
        & = - i [ {\mathcal H}_c + {\mathcal H}_q, \rho ] .
    \end{align}
    The classical system can be represented using the classical master equation,
    \begin{equation}
        \frac{d P_i}{d t} = \sum_j {\mathcal L}_{ij} P_j ,
    \end{equation}
    where $P_j$ is the probability of the $j$-th state, and ${\mathcal L}_{ij}$ is the transition rate matrix.
    For an Ising spin system, this matrix is defined in terms of single-spin flips,
    \begin{equation}
        {\mathcal L}_{ij} = \begin{cases}
            \frac{e^{-\beta E_i}}{e^{-\beta E_i}+e^{-\beta E_j}} , & \text{single-spin flip} , \\
            - \sum_{k\neq i} {\mathcal L}_{ki} ,                   & i = j ,\\
            0 ,                                                    & \text{otherwise} .
        \end{cases}
    \end{equation}
    The parameter $\beta$ is the inverse temperature ($= 1/T$), whereas $E_i$ is the energy of the $i$-th state of the diagonal Hamiltonian ${\mathcal H}_c$.

    To interpolate the two dynamics, we do not modify the differential equations; however, we continuously intervene in their solutions.
    We use a hierarchical structure by considering the two types of dynamics as lower-level processes to be integrated.
    Not only is this formulation easy to be implemented in numerical simulations, but it also enables a single-layered expression, i.e., a set of differential equations to be obtained once the Hamiltonian has been provided.
    We only investigate this differential-equation representation for a two-level system in this study.
    However, similar calculations could be performed in principle for any system.

    The actual process used to interpolate the two dynamics is as follows.
    Each system evolves from the same initial state (typically an equal superposition of all possible states) using their respective dynamics.
    After each infinitesimal time step $dt$, we construct a new density matrix $\tilde{\rho}$ and a new probability vector $\tilde{P}$ using the previous density matrix $\rho$ (from the von Neumann equation) and probability vector $P$ (from the classical master equation):
    \begin{equation}
        \label{eq_rule}
        \tilde{\rho}(t+dt) = \begin{pmatrix}
            r_1^2                   & r_1 r_2 \hat{\rho}_{12} & \cdots & r_1 r_N \hat{\rho}_{1N} \\
            r_2 r_1 \hat{\rho}_{21} & r_2^2                   & \cdots & r_2 r_N \hat{\rho}_{2N} \\
            \vdots                  & \vdots                  & \ddots & \vdots \\
            r_N r_1 \hat{\rho}_{N1} & r_N r_2 \hat{\rho}_{N2} & \cdots & r_N^2 \\
        \end{pmatrix}
    \end{equation}
    and
    \begin{equation}
        \tilde{P}(t+dt) = ( r_1^2 \ \ r_2^2 \ \ \cdots \ \ r_N^2 ) ,
    \end{equation}
    where
    \begin{equation}
        \hat{\rho}_{ij} = \frac{\rho_{ij}(t+dt)}{|\rho_{ij}(t+dt)|}
    \end{equation}
    and
    \begin{equation}
        r_i = \sqrt{(1-\alpha)\rho_{ii}(t+dt) + \alpha P_i(t+dt)} .
    \end{equation}
    The new $\tilde{\rho}(t+dt)$ and $\tilde{P}(t+dt)$ further become the density matrix and probability vector for the succeeding time step.
    The total system consists of two subsystems, which are governed by different dynamics, that influence each other using the aforementioned interpolation process.

    This definition indicates that the density matrix satisfies the pure state condition,
    \begin{equation}
        \label{eq_pure_state_condition}
        \tilde{\rho}^2 = \tilde{\rho} .
    \end{equation}
    As the system is in a pure state, it can be equivalently updated using the state vector representation of the Schr\"odinger equation.
    Let the state vector $\ket{\psi}$ be a linear combination of the eigenvectors:
    \begin{equation}
        \ket{\psi(t)} = \sum_i a_i(t) \ket{i} .
    \end{equation}
    Thus, the update rules can be written as
    \begin{align}
        & \ket{\tilde{\psi}(t+dt)} = \nonumber \\
        & \sum_i \sqrt{(1-\alpha) |a_i(t+dt)|^2 + \alpha P_i(t+dt)} \frac{a_i(t+dt)}{|a_i(t+dt)|} \ket{i}
    \end{align}
    and
    \begin{equation}
        \tilde{P}_i(t+dt) = (1-\alpha) |a_i(t+dt)|^2 + \alpha P_i(t+dt) .
    \end{equation}

    \section{Two-level system}
    \label{sec_tls}
    In this section, we investigate the mixed dynamics using a simple two-level system, with the Hamiltonian
    \begin{align}
        {\mathcal H}_c & = - h \sigma^z , \\
        {\mathcal H}_q & = - \Gamma \sigma^x ,
    \end{align}
    and
    \begin{equation}
        {\mathcal H} = {\mathcal H}_c + {\mathcal H}_q =
        \begin{pmatrix}
            - h      & - \Gamma \\
            - \Gamma & h
        \end{pmatrix} ,
    \end{equation}
    where $h$ and $\Gamma$ are positive.
    Substituting this Hamiltonian and the explicit form of the density matrix,
    \begin{equation}
        \rho = \begin{pmatrix}
            \rho_{11} & \rho_{12} \\
            \rho_{21} & \rho_{22}
        \end{pmatrix} ,
    \end{equation}
    into the von Neumann equation, we obtain
    \begin{align}
        \label{eq_tls_vne}
        & \begin{pmatrix}
            \dot{\rho}_{11} & \dot{\rho}_{12} \\
            \dot{\rho}_{21} & \dot{\rho}_{22}
        \end{pmatrix} = \nonumber \\
        & \begin{pmatrix}
        - i \Gamma (\rho_{12} - \rho_{21})                 & - i \Gamma (\rho_{11} - \rho_{22}) + 2 i h \rho_{12} \\
            i \Gamma (\rho_{11} - \rho_{22}) - 2 i h \rho_{21} & i \Gamma (\rho_{12} - \rho_{21})
        \end{pmatrix} .
    \end{align}
    The classical master equation can be given as
    \begin{equation}
        \begin{pmatrix}
            \dot{P}_{1} \\
            \dot{P}_{2}
        \end{pmatrix} = \frac{1}{e^{\beta h} + e^{-\beta h}} \begin{pmatrix}
            - e^{-\beta h} P_{1} + e^{\beta h} P_{2} \\
            e^{-\beta h} P_{1} - e^{\beta h} P_{2}
        \end{pmatrix} .
    \end{equation}
    To ensure simplicity, we initially consider the zero temperature limit, $T=0$ ($\beta = \infty$).
    In this situation, the equation takes the simple form as follows:
    \begin{equation}
        \label{eq_tls_me}
        \begin{pmatrix}
            \dot{P}_{1} \\
            \dot{P}_{2}
        \end{pmatrix} = \begin{pmatrix}
            P_{2} \\
            - P_{2}
        \end{pmatrix} .
    \end{equation}
    Substituting Eqs.~(\ref{eq_tls_vne}) and (\ref{eq_tls_me}) into Eq.~(\ref{eq_rule}), we obtain the following:
    \begin{align}
        \label{eq_r11}
        \dot{\rho}_{11} & = - i (1 - \alpha) \Gamma (\rho_{12} - \rho_{21}) + \alpha \rho_{22} , \\
        \label{eq_r12}
        \dot{\rho}_{12} & = - i (1 - b \alpha) \Gamma (\rho_{11} - \rho_{22}) - (c \frac{\alpha}{2} - 2 i h) \rho_{12} ,
    \end{align}
    where
    \begin{align}
        \label{eq_b_zero_T}
        b & \equiv \frac{1}{2} \left[ 1 - \left( \frac{\rho_{12}}{|\rho_{12}|} \right)^2 \right] \;\;\;\;\; \text{($0 \le |b| \le 1$)} , \\
        \label{eq_c_zero_T}
        c & \equiv \frac{(\rho_{11} - \rho_{22})}{\rho_{11}} .
    \end{align}
    A detailed derivation of this is in Appendix~\ref{sec_tls_DEs}.

\begin{figure}[thb]
        \includegraphics[width=80mm]{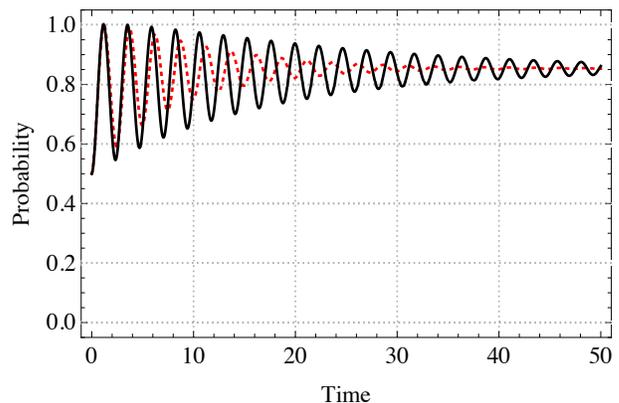}%
        \caption{
            \label{fig_TLS_dynamics}
            Time evolution of the probability $\rho_{11}$ with $T = 0$ and $h = \Gamma = 1$; for $\alpha = 0.1$ (solid black) and $0.2$ (dashed red).
        }
    \end{figure}

    The dynamics of Eqs.~(\ref{eq_r11}) and (\ref{eq_r12}) are depicted in Fig.~\ref{fig_TLS_dynamics}.
    The two lines are associated with different values of the mixing parameter $\alpha$, which illustrates that the parameter controls the relaxation time of the system as expected.

    These equations are similar to the optical Bloch equations\cite{Metcalf1999,Arecchi1965}, which were observed to be as follows:
    \begin{align}
        \label{eq_rgg}
        \frac{d\rho_{gg}}{dt} & = + \gamma \rho_{ee} + \frac{i}{2}(\Omega^* {\tilde \rho_{eg}} - \Omega {\tilde \rho_{ge}}) , \\
        \label{eq_rge}
        \frac{d{\tilde \rho_{ge}}}{dt} & = - (\frac{\gamma}{2} + i\delta){\tilde \rho_{ge}} + \frac{i}{2} \Omega^* (\rho_{ee} - \rho_{gg}) .
    \end{align}
    However, the mixed-dynamics system is in a pure state while the optical Bloch system is in a mixed state.
    This is because Eqs.~(\ref{eq_rgg}) and (\ref{eq_rge}) are linear but Eqs.~(\ref{eq_r11}) and (\ref{eq_r12}) are nonlinear, and these nonlinear equations enable the system to maintain a pure state.
    In the case of $\alpha \ll 1$ and $\rho_{22} \ll \rho_{11}$, which represent the weak coupling and asymptotic limits, they are reduced to linear equations and become equivalent to the open Bloch equations.

    We can obtain the steady state of Eqs.~(\ref{eq_r11}) and (\ref{eq_r12}) at zero temperature by splitting the complex numbers into real and imaginary parts as $\rho_{12}=x+iy$ and $\rho_{11}=z$, which gives
    \begin{align}
        \alpha \Gamma \frac{x y (2 z - 1)}{x^2 + y^2} - \alpha \frac{x (2 z - 1)}{2 z} - 2 h y & = 0 , \\
        \Gamma \frac{[- x^2 - (1 - \alpha) y^2] (2 z - 1)}{x^2 + y^2} & \nonumber \\
        - \alpha \frac{y (2 z - 1)}{2 z} + 2 h x & = 0 , \\
        2 (1 - \alpha) \Gamma y + \alpha (1 - z) & = 0 .
    \end{align}

\begin{figure}[thb]
        \includegraphics[width=80mm]{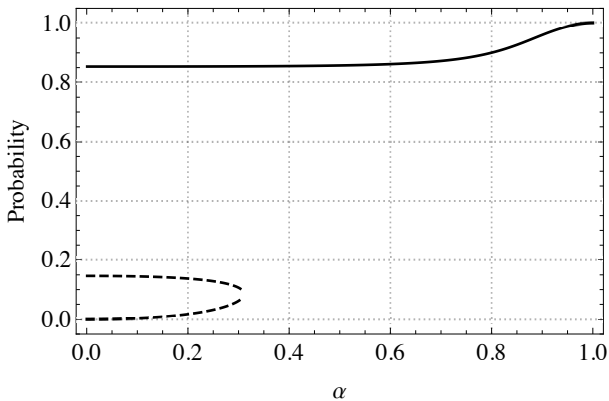}%
        \caption{
            \label{fig_TLS_T0}
            Stationary-state probabilities $\rho_{11}$ as a function of $\alpha$ for $T = 0$ and $h = \Gamma = 1$.
            The solid and dashed lines represent the stable and unstable solutions, respectively.
        }
    \end{figure}

    Figure~\ref{fig_TLS_T0} depicts the probability $\rho_{11} \ (=z)$ as a function of $\alpha$.
    When $\alpha = 0$, the stable solution converges to the ground-state probability of the given quantum Hamiltonian ${\mathcal H} (= {\mathcal H}_c + {\mathcal H}_q)$, which is $(2+\sqrt{2})/4 \ (\sim 0.854)$ for $h=\Gamma=1$.
    It recovers the probability of the classical Hamiltonian ${\mathcal H}_c$ (i.e., $\rho_{11} = 1$) when $\alpha = 1$.
    Further, it interpolates smoothly and monotonically between these two extremes.
    Additionally, the system remains close to the ground state of the quantum Hamiltonian over an extensive range of $\alpha$, implying that the quantum dynamics are stable against thermal intervention using the classical master equation.

    The remainder of this section analyzes the behavior of the mixed dynamics at finite temperatures.
    In this situation, Eqs.~(\ref{eq_r11}) and (\ref{eq_c_zero_T}) become
    \begin{equation}
        \dot{\rho}_{11} = - i (1 - \alpha) \Gamma (\rho_{12} - \rho_{21}) + \alpha \frac{(\rho_{22} e^{\beta h} - \rho_{11} e^{-\beta h})}{(e^{\beta h} + e^{-\beta h})}
    \end{equation}
    and
    \begin{equation}
        c \equiv \frac{(\rho_{11} - \rho_{22}) (\rho_{22} e^{\beta h} - \rho_{11} e^{-\beta h})}{\rho_{11} \rho_{22} (e^{\beta h} + e^{-\beta h})} .
    \end{equation}

\begin{figure}[thb]
        \includegraphics[width=80mm]{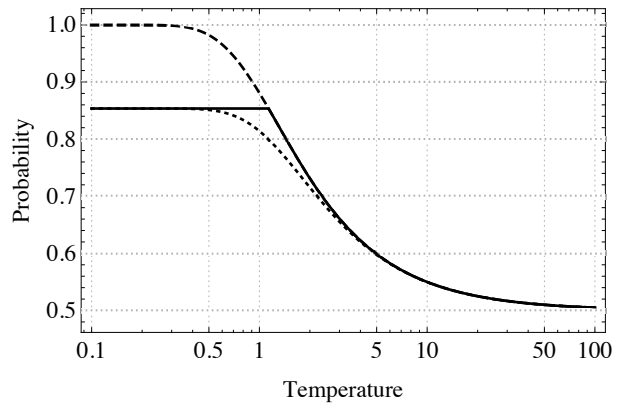}%
        \caption{
            \label{fig_TLS_Tdep}
            Stable stationary-state probability $\rho_{11}$ as a function of $T$, with $h = \Gamma = 1$ and $\alpha = 0.1$ (solid line).
            The dashed and dotted lines represent the probabilities derived from the standard partition function of the total Hamiltonian with $\Gamma = 0$ and $1$.
        }
    \end{figure}

    The finite temperature results are depicted in Fig.~\ref{fig_TLS_T02to100} in Appendix~\ref{sec_figs}.
    As expected, the probability at $\alpha = 1$ is identical to that provided by the classical master equation at all temperatures.
    However, the solution is nontrivial at lower values of $\alpha$.
    Figure~\ref{fig_TLS_Tdep} depicts the temperature dependency of the probability at $\alpha = 0.1$ along with the associated ground-state probabilities for the classical and quantum Hamiltonians, ${\mathcal H}_c$ and $\mathcal H$.
    Below a temperature of approximately $T = 1.1$, the system remains in the quantum ground state; however, it begins to behave like a classical system after the intersection point with the classical Hamiltonian curve.
    This intersection point remains at the limit of $\alpha \to 0$.

    In the two-level system, the ground state of the quantum Hamiltonian is stable against intervention by the classical master equation for small values of both the mixing parameter $\alpha$ and the temperature $T$.
    This is a potentially useful characteristic for finding the ground states of the quantum systems by calculating their relaxation processes over time instead of directly diagonalizing the Hamiltonian.
    However, further investigation will be required for such applications because our study only provides results for the two-level (single-spin) system.

    \section{Numerical Simulations}
    \label{sec_simulation}
    In this section, we perform numerical simulations of the multispin systems.
    To ensure simplicity, all the simulations were conducted at zero temperature.
    Since we intended to investigate the dynamics of QA, we use the Ising Hamiltonian formulation, which is compatible with the QUBO problems.
    Although we use a transverse field as a quantum driving force to tunnel between various states, other quantum sources, such as nonstoquastic Hamiltonians\cite{Farhi2002,Seki2012,Seoane2012,Crosson2014,Seki2015,Hormozi2017,Nishimori2017}, could also be used.
    The general form of this Hamiltonian is
    \begin{align}
        {\mathcal H}   & = s {\mathcal H}_c + (1-s) {\mathcal H}_q , \\
        {\mathcal H}_c & = - \sum_{(ij)} J_{ij} \sigma_i^z \sigma_j^z - \sum_i h_i \sigma_i^z , \\
        {\mathcal H}_q & = - \sum_i \sigma_i^x ,
    \end{align}
    where ${\mathcal H}_c$ represents the QUBO Hamiltonian whose ground state has to be obtained, and ${\mathcal H}_q$ represents the transverse field.
    The ground state of the total Hamiltonian is controlled by the parameter $s$, which increases monotonically over time: a system starts from a superposition of all possible spin configurations at $s = 0$ and converges on the ground state of the QUBO Hamiltonian at $s = 1$.
    As $s$ evolves from 0 to 1, the ground state of the total Hamiltonian alters from a trivial ground state to a nontrivial QUBO solution.

\begin{figure}[thb]
        \includegraphics[width=80mm]{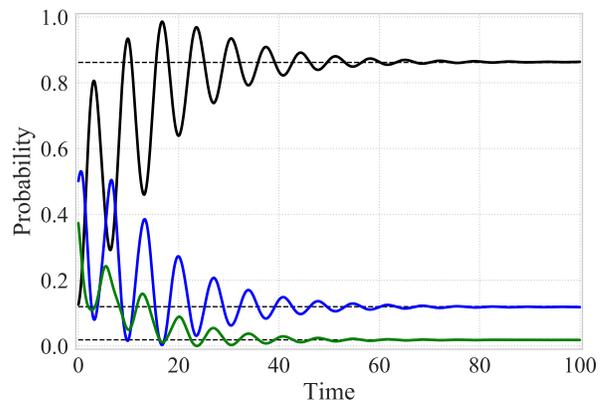}%
        \caption{
            \label{fig_HT_quench}
            Time evolution of the total probabilities for the states associated with the three energy levels of the HT model for $\alpha = 0.1$, $T = 0$, and $s = 0.8$.
            Going from top to bottom, the solid lines represent the ground state (black) and the first (blue) and second (green) excited states of the QUBO Hamiltonian, while the dashed lines represent the probabilities of the stationary state.
        }
    \end{figure}

    The first example we consider is a 4-spin Husimi--Temperley (HT) model, i.e., an infinite-range ferromagnetic model.
    This model can be expressed as $J_{ij} = 1/4$ and $h_i = 0$, with all 16 states falling into one of the three degenerate energy levels.
    Figure~\ref{fig_HT_quench} depicts the quenched dynamics at $\alpha = 0.1$, $T = 0$, and $s = 0.8$ in the presence of quantum effects.
    Although $s$ should be increased as a function of $t$ in standard QA simulations, we maintained the parameters to be constant in order to investigate the relaxation process.
    The ground state of the total Hamiltonian retains $\sigma^x$ and $\sigma^y$ components, indicating that it overlaps with both the ground state of the QUBO Hamiltonian and its two excited states.
    The three solid lines represent the probabilities of these states in the simulation, while the dashed lines represent the stationary states of the total Hamiltonian.
    As depicted in the previous example, the system appears to relax to the ground state.

\begin{figure}[thb]
        \includegraphics[width=80mm]{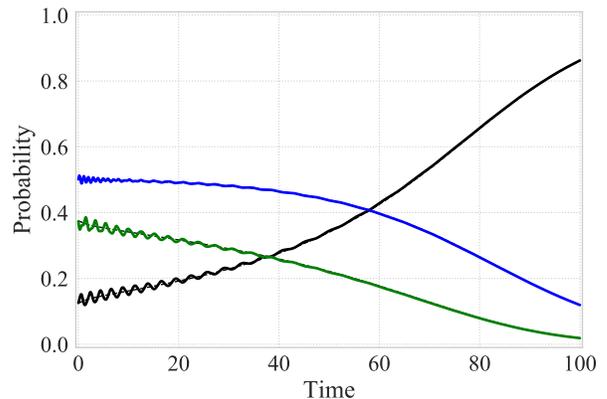}%
        \caption{
            \label{fig_HT_anneal}
            Time evolution of the total probabilities for the states associated with the three energy levels of the HT model for $\alpha = 0.1$, $T = 0$, and $s(t) =  0.8 \sqrt{t/100}$.
            From top to bottom at the end of the simulation, the solid lines represent the ground state (black) and the first (blue) and second (green) excited states of the QUBO Hamiltonian, while the dashed lines represent the stationary-state probabilities.
            }
    \end{figure}

    We further conducted a standard QA simulation where $s$ increased over time according to $s(t) = 0.8 \sqrt{t/100}$, and the results are depicted in Fig.~\ref{fig_HT_anneal}.
    It was observed that the thermal effects from the classical master equation drove the system during the initial simulation; however, both the thermal and quantum effects drive the system during this simulation.
    These results depict less oscillation around the stationary states because the larger oscillations during the initial simulation were caused by the sudden alteration of the parameter $s$ from 0 (at $t = 0$) to 0.8 (for $t > 0$).
    Thus, the system remains in a stationary state even in the case of the mixed dynamics if the annealing schedule is selected appropriately.

    Further, we investigate the intermediate dynamics (controlled by the mixing parameter $\alpha$) by concentrating on a system known to show clear differences between its quantum and classical dynamics.
    The quantum signature (QS) Hamiltonian was designed to help distinguish between simulated annealing (SA) and QA\cite{Boixo2013,Albash2015}.
    The eight-spin version of this Hamiltonian consists of a four-spin ring of ``core spins'' and four other ``outer spins,'' each attached to a different core spin, with all the ferromagnetic interactions taking the form $J_{ij} = 1$.
    The local fields for the core and outer spins are $h_i = 1$ and $-1$, respectively.
    Due to the frustration caused by the competition between the ferromagnetic interaction and local field, the ground states are observed to degenerate and can be divided into two categories: a cluster of 16 ground states with four up-spins in the core and an isolated ground state with all eight spins being down, as follows:
    \begin{align}
        C & = \{ \; \ket{\uparrow \uparrow \uparrow \uparrow \ \updownarrow \updownarrow \updownarrow \updownarrow} \, \} \;\;\; \text{(cluster of 16 states)} \nonumber ,\\
        I & = \{ \; \ket{\downarrow \downarrow \downarrow \downarrow \ \downarrow \downarrow \downarrow \downarrow} \, \} \;\;\; \text{(isolated state)} \nonumber .
    \end{align}

\begin{figure}[thb]
        \includegraphics[width=80mm]{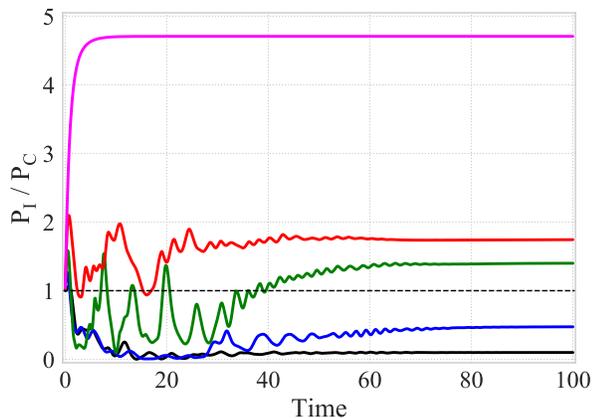}%
        \caption{
            \label{fig_QS_pipc}
            Time evolution of the 8-spin QS system's $P_I / P_C$ ratio for (from bottom to top) $\alpha = 0$ (black), 0.25 (blue), 0.5 (green), 0.75 (red), and 1 (purple).
        }
    \end{figure}

    During the annealing process, the average clustered state probability $P_C = 1 / 16 \sum_{c \in C} P_c$ develops faster than the isolated state probability $P_I$ in QA ($P_I / P_C < 1$).
    Oppositely, the isolated state becomes dominant in SA ($P_I / P_C > 1$)\cite{Boixo2013,Albash2015}.
    The time evolution of the $P_I / P_C$ ratio is depicted for $\alpha = 0, 0.25, 0.5, 0.75$, and $1$ in Fig.~\ref{fig_QS_pipc}.
    In this simulation, the quantum annealing scheduled was $s(t) = \sqrt{t/100}$, while the temperature was not annealed but instead quenched to $T = 0$, which represented the limit of the fastest possible annealing schedule.
    Even though the temperature was not annealed, we still invoke this classical master-equation-driven dynamics SA to imply that the quenched temperature could be modified to an annealed temperature without causing any significant changes.
    The results of QA and SA ($\alpha = 0$ and $1$) depict the lowest and highest $P_I / P_C$ ratios at the end of the simulation, respectively, whereas the mixed results fall between these values in accordance with the amplitude of $\alpha$.

    Finally, we consider the SK model, which introduces randomness into the interactions $J_{ij}$.
    These interactions obey the Gaussian distribution $J_{ij} \sim {\mathcal N}(0, 1/N)$, where $N$ is the size of the system, and the local field is set to zero ($h_i = 0$).
    The energy landscape of the SK model is complex, which includes $2^{N-1}$ different energy levels that are defined by the random interactions.
    To investigate the general behavior of the dynamics of this model, numerical simulations were conducted for 50 different sets of interactions, and the ground-state probabilities were averaged.
    We investigated the dependence of these probabilities on the total annealing time $\tau$ and mixing parameter $\alpha$.
    We quenched the temperature to $T = 0$ as the operation temperature of the real annealing machines is observed to be low enough in order to make this a reasonable approximation.
    As the optimal schedule was unknown, we explored annealing schedules of the form $s(t) = (t / \tau)^\gamma$, where $\gamma$ is a scheduling index that controls the shape of the scheduling curve.

\begin{figure}[thb]
        \includegraphics[width=80mm]{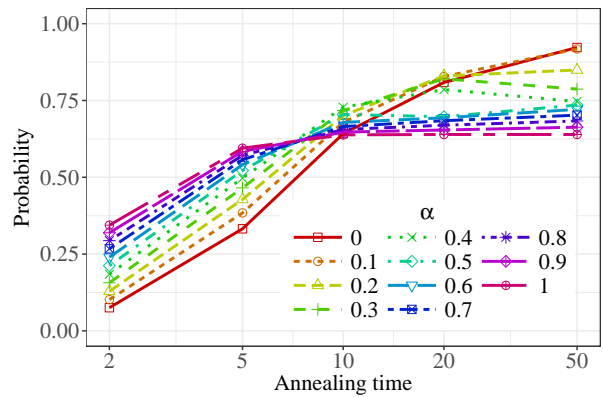}%
        \caption{
            \label{fig_SK_qeff04}
            QUBO ground-state probability as a function of annealing time for the mixing parameters $\alpha = 0, 0.1, 0.2, \dots, 1$ under the annealing schedule $s(t) = (t / \tau)^{0.4}$.
        }
    \end{figure}

    Figure \ref{fig_SK_qeff04} depicts the results for the specific schedule $s(t) = (t / \tau)^{0.4}$, while Fig.~\ref{fig_SK_t02to50} illustrates all the tested annealing schedules.
    For the shortest annealing time, $\tau = 2$, the classical system ($\alpha = 1$) depicted the optimal performance.
    This indicates that the annealing schedule was too rapid for the quantum system to follow its instantaneous ground state, while thermal effects allowed the classical system to relax to lower energy states.
    However, QA depicted the optimal performance for the longest annealing time, $\tau = 50$, indicating that it was able to follow its instantaneous ground state\cite{Kadowaki1998}.
    QA can be effective in such situations.
    For intermediate annealing times, the best performance was achieved at intermediate $\alpha$ values, such as $\alpha = 0.4$ and $0.2$ for $\tau = 10$ and $20$, respectively.
    These results are nontrivial, but can be understood from the fact that the best dynamics changes from SA to QA according to annealing times, implying the possibility that mixed dynamics could outperform both at intermediate annealing times.
    Adding thermal fluctuations to quantum systems can result in a better performance than either QA or SA in certain situations.

    \section{Summary and Discussion}
    \label{sec_summary}
    In this study, we have proposed a method for interpolating between von Neumann and classical master equations to introduce thermal effects into quantum dynamics.
    These intermediate dynamics bridge the two equations using the mixing parameter $\alpha$.

    For the two-level system, these dynamics can be represented using differential equations.
    These are equivalent to the optical Bloch equations\cite{Arecchi1965} at the weak coupling and asymptotic limits, wihch implies a connection between our approach and the phenomenological equations.
    Relaxation to a stationary state is governed by the mixing parameter and temperature that can be considered to be the system-bath coupling constant and bath temperature, respectively.
    Stationary analysis of this model revealed that the solution was smoothly altered from quantum to classical as a function of $\alpha$; however, it is not differentiable at a certain point when expressed as a function of temperature.
    If this non-differentiable point is also present for multi-spin systems, it may potentially be associated with a phase transition in the two-dimensional parameter space spanning $T$ and $\alpha$.
    This assumption will be investigated further in future work.
    Additionally, the resulting stationary state represents the ground state of the quantum Hamiltonian in a certain parameter-space region where $T$ and $\alpha$ are both low.
    If this additionally holds true for many-body systems, the proposed dynamics could be an alternative method to estimate their ground states.

    The results of numerically simulating an example many-body system (the four-spin HT model) confirmed that it relaxes to a stationary state similar to that observed in the two-level system, indicating that QA still works when thermal fluctuations are introduced into the quantum dynamics using the classical master equation.
    Although this system was still small, the fact that a similar relaxation process was observed implies that the proposed dynamics can naturally interpolate quantum dynamics and thermodynamics in many-body systems.

    An analysis of the QS model provided detailed information about the behavior of the interpolated dynamics because the ground-state probabilities of the model reflect both the quantum and classical dynamics.
    These results clearly demonstrated that increasing the mixing parameter $\alpha$ resulted in the system behavior becoming approximately similar to that of pure thermodynamics.
    Although a classical $O(2)$ spin model can provide a good approximation of QA in certain circumstances\cite{Shin2014}, our analysis did not use the $P_I / P_C$ ratio of the model to determine whether a given ``black box'' was quantum or classical; however, we used the corresponding ratio to demonstrate that the overlap between the two types of dynamics could be controlled by the mixing parameter.
    While the former topic has been extensively discussed\cite{Boixo2013,Shin2014,Smolin2014,Albash2015}, it is not directly linked to our objective in this study.

    Finally, we tested the SK model, a spin glass model, as an example of a typical combinatorial optimization problem.
    We observed that SA depicted a higher probability of finding the ground state for short annealing times, while QA performed better for long annealing times.
    From this observation, it was clear that QA tends to follow its instantaneous ground state with an appropriate (longer) annealing schedule even though SA can work with quenching (short annealing time) by thermal relaxation\cite{Kadowaki1998}.
    For annealing times where SA and QA produced comparable performances, the proposed mixed dynamics demonstrated better performance than either of them.
    In such cases, quantum and thermal fluctuations may work together, as several studies have reported\cite{Amin2008,Venuti2017,Dickson2013,Nishimura2016}, in order to improve the performance.
    This observation may suggest methods of selecting optimal annealing times and schedules to increase the probability of recovering ground-state solutions with real quantum annealers.

    Before applying the proposed dynamics, we should gain a broader understanding about them.
    This is why we noted their similarity to the optical Bloch equations in certain situations.
    One difference highlighted by this comparison is that the proposed dynamics maintains the system in a pure state.
    In the weak coupling and asymptotic limits, the dynamics are reduced to linear equations that are equivalent to the open Bloch equations.
    Systems that maintain pure states have also been investigated by different methods such as the eigenstate thermalization hypothesis (ETH) and generalized Gibbs ensemble (GGE)\cite{Srednicki1994,DAlessio2016}.
    Although our approach is similar to them with regard to whether the system maintains a pure or mixed state, they consider isolated quantum systems that relax to equilibrium states but do not dissipate energy.
    A comparison with these approaches may be another path to understanding the proposed dynamics extensively.

    Regarding the system-bath coupling, our formulation does not allow a microscopic coupling mode to be directly specified.
    In contrast with the quantum fluctuations caused by the transverse field coupling with $\sigma^x$, thermal relaxation occurs via $\sigma^z$ as the Hamiltonian in the classical master equation is represented by $\sigma^z$.
    However, the coupling can be specified indirectly by selecting different classical and quantum Hamiltonians for the classical master and von Neumann equations.
    Using different Hamiltonians for the quantum and classical dynamics would allow us to design specific couplings that can be used to model real devices.

    System size is a limiting factor while using our proposed dynamics.
    Since the dimension of the matrix in the calculation scales as $2^N$, relatively small systems could tax the available computational resources.
    Handling larger systems would require the use of Monte Carlo simulations; however, it is not obvious how such simulations can be conducted.
    Another limitation of the current study is that most of the simulations were conducted at zero temperature.
    While the results for the two-level system depicted that the zero-temperature solution remained stable at low temperatures, we have to confirm the robustness of quantum dynamics of many-body systems at finite temperatures.
    The behavior of open quantum systems can be parameterized by the environment temperature and coupling constant, and the two-level system demonstrated nontrivial behavior as a function of these parameters (Figs.~\ref{fig_TLS_Tdep} and \ref{fig_TLS_T02to100}).
    Further, the quantum ground state did not alter at low temperatures; however, their temperature stability should also be investigated in multi-spin systems.
    Finally, our most essential future study will involve the comparison of our numerical simulation results with the output of a real annealer to depict the accuracy of prediction of the experimental results using the proposed dynamics.
    As Albash \textit{et al.} pointed out\cite{Albash2017}, being able to predict the behavior of a quantum annealer at finite temperatures will enable us to design devices with larger numbers of qubits.

    \section*{Acknowledgments}
    We are grateful to H. Nishimori, S. Miyashita, S. Tanaka, M. Ohzeki and M. Okuyama for their helpful discussions and suggestions.

    \begin{widetext}
    \appendix
    \section{Differential equations for the two-level system}
    \label{sec_tls_DEs}
    The updated density matrix elements at $t+dt$ are as follows:
    \begin{align}
        \label{eq_a_r11}
        \tilde{\rho}_{11}
        & = (1 - \alpha) [\rho_{11} - i \Gamma (\rho_{12} - \rho_{21}) dt]
        + \alpha (\rho_{11} + \rho_{22} dt) \nonumber , \\
        & = \rho_{11} - i (1 - \alpha) \Gamma (\rho_{12} - \rho_{21}) dt + \alpha \rho_{22} dt ,
    \end{align}
    \begin{equation}
        \label{eq_a_r22}
        \tilde{\rho}_{22} = \rho_{22} + i (1 - \alpha) \Gamma (\rho_{12} - \rho_{21}) dt - \alpha \rho_{22} dt
    \end{equation}
    \begin{align}
        \lefteqn{\tilde{\rho}_{12} = r_1 r_2 \frac{\rho_{12}}{|\rho_{12}|}} \\
        & = \{ [ \rho_{11} - i (1 - \alpha) \Gamma (\rho_{12} - \rho_{21}) dt + \alpha \rho_{22} dt ]
        [ \rho_{22} + i (1 - \alpha) \Gamma (\rho_{12} - \rho_{21}) dt - \alpha \rho_{22} dt ] \}^{\frac{1}{2}} \nonumber \\
        & \quad \times \frac{\rho_{12} - i \Gamma (\rho_{11} - \rho_{22}) dt + 2 i h \rho_{12} dt}
        {\{ [ \rho_{12} - i \Gamma (\rho_{11} - \rho_{22}) dt + 2 i h \rho_{12} dt ]
        [ \rho^*_{12} + i \Gamma (\rho_{11} - \rho_{22}) dt - 2 i h \rho^*_{12} dt ] \}^{\frac{1}{2}}} \\
        & = \sqrt{\rho_{11} \rho_{22}} \left[ 1 + i (1 - \alpha) \Gamma \frac{(\rho_{11} - \rho_{22}) (\rho_{12} - \rho_{21})}{\rho_{11} \rho_{22}} dt - \alpha \frac{(\rho_{11} - \rho_{22})}{\rho_{11}} dt + O(dt^2) \right]^{\frac{1}{2}} \nonumber \\
        & \quad \times \rho_{12} \left[ 1 - i \Gamma \frac{(\rho_{11} - \rho_{22})}{\rho_{12}} + 2 i h dt \right] \frac{1}{\sqrt{\rho_{12} \rho^*_{12}}} \left[ 1 + i \Gamma \frac{(\rho_{11} - \rho_{22}) (\rho_{12} - \rho^*_{12})}{\rho_{12} \rho^*_{12}} dt + O(dt^2) \right]^{-\frac{1}{2}} \\
        & = \rho_{12} \sqrt{\frac{\rho_{11} \rho_{22}}{\rho_{12} \rho_{21}}} \left\{ 1 + \frac{i \Gamma (\rho_{11} - \rho_{22})}{2} \left[ \frac{(1 - \alpha) (\rho_{12} - \rho_{21})}{\rho_{11} \rho_{22}} - \frac{(\rho_{12} + \rho_{21})}{\rho_{12} \rho_{21}} \right] dt + 2 i h dt - \frac{(\rho_{11} - \rho_{22})}{\rho_{11}} \frac{\alpha}{2} dt + O(dt^2) \right\} \\
        \label{eq_a_r12}
        & = \rho_{12} - i \left\{ 1 - \frac{1}{2} \left[ 1 - \left( \frac{\rho_{12}}{|\rho_{12}|} \right)^2 \right] \alpha \right\} \Gamma (\rho_{11} - \rho_{22}) dt + 2 i h \rho_{12} dt - \frac{(\rho_{11} - \rho_{22})}{\rho_{11}} \frac{\alpha}{2} \rho_{12} dt + O(dt^2) .
    \end{align}
    Here, we have used $P_i = \rho_{ii}$, $\rho^*_{12} = \rho_{21}$, and $\rho_{11} \rho_{22} = \rho_{12} \rho_{21}$ [derived from Eq.~(\ref{eq_pure_state_condition})].
    Additionally, the $\tilde{\rho}_{ij}(t+dt)$ values from equations (\ref{eq_a_r11}), (\ref{eq_a_r22}), and (\ref{eq_a_r12}) are used to substitute for $\rho_{ij}(t+dt)$ while calculating the next time step (from $t+dt$ to $t+2dt$) of the quantum and classical dynamics.
    These equations can therefore be regarded as representing the dynamics of the mixed system, and differential equation representation can be obtained from them.

    \section{Additional figures}
    \label{sec_figs}
    \begin{figure*}[htb]
        \centering
        \begin{tabular}{l}
            \begin{minipage}{0.33\textwidth}
                \includegraphics[width=60.63mm,clip,trim=0 13 0 0]{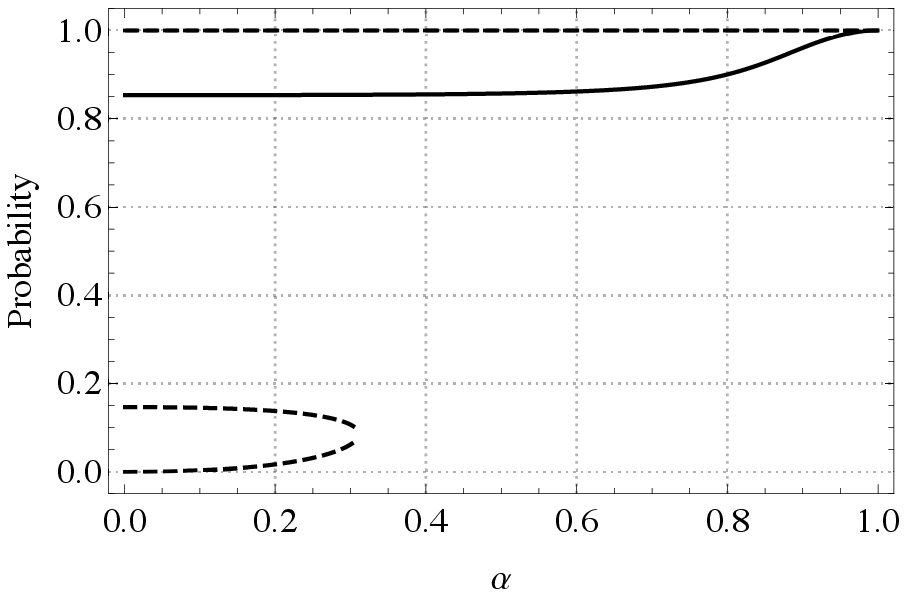}%
            \end{minipage}
            \begin{minipage}{0.31465\textwidth}
                \includegraphics[width=57.81mm,clip,trim=13 13 0 0]{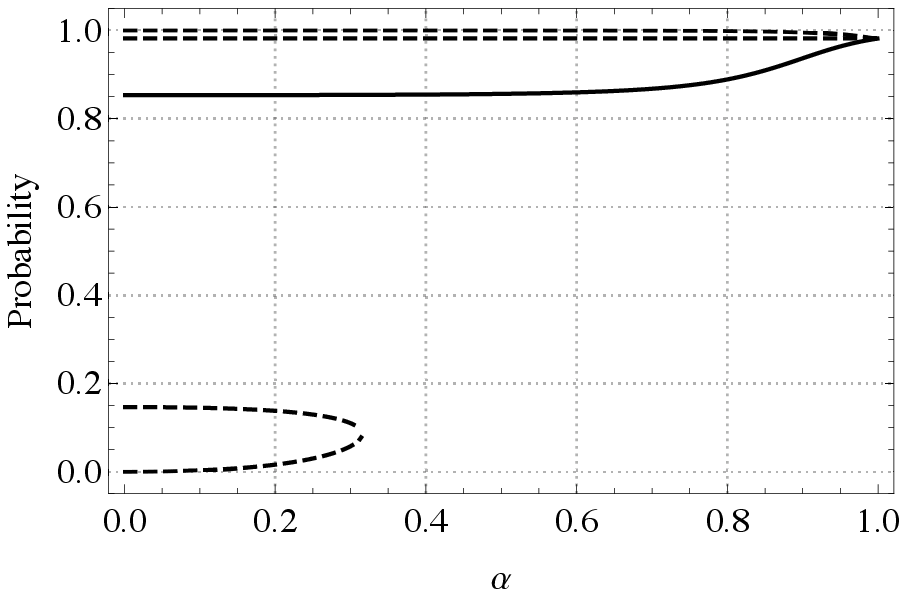}%
            \end{minipage}
            \begin{minipage}{0.31465\textwidth}
                \includegraphics[width=57.81mm,clip,trim=13 13 0 0]{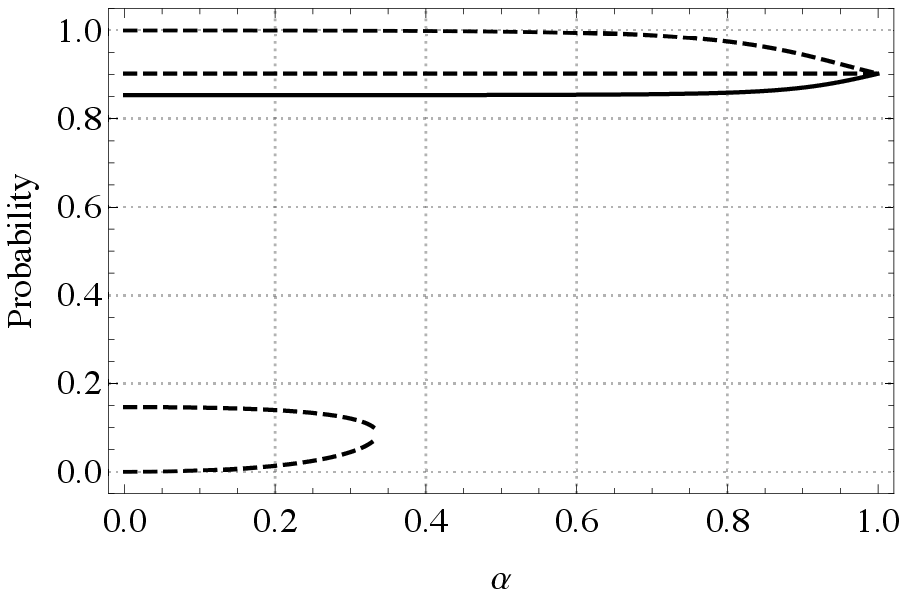}%
            \end{minipage} \\
            \begin{minipage}{0.33\textwidth}
                \includegraphics[width=60.63mm,clip,trim=0 13 0 0]{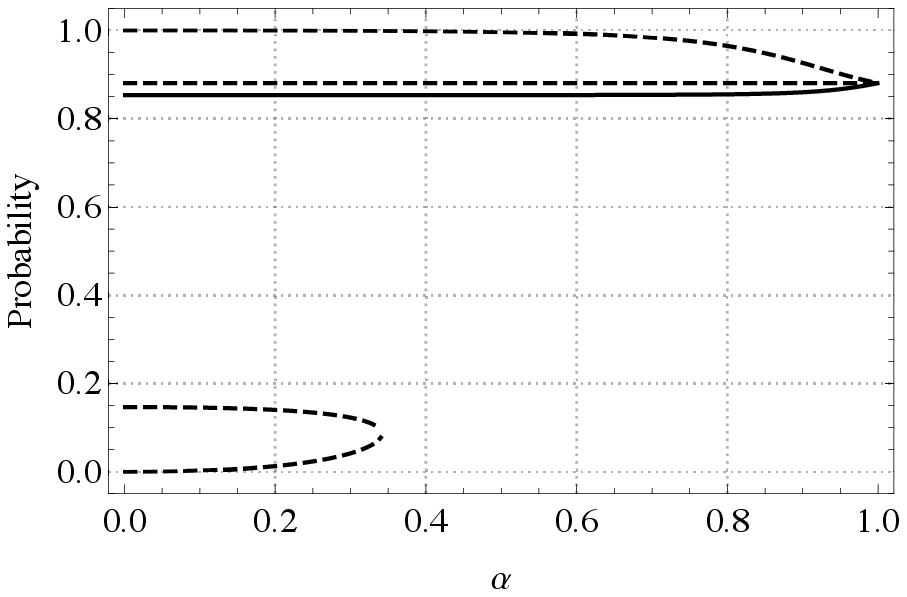}%
            \end{minipage}
            \begin{minipage}{0.31465\textwidth}
                \includegraphics[width=57.81mm,clip,trim=13 13 0 0]{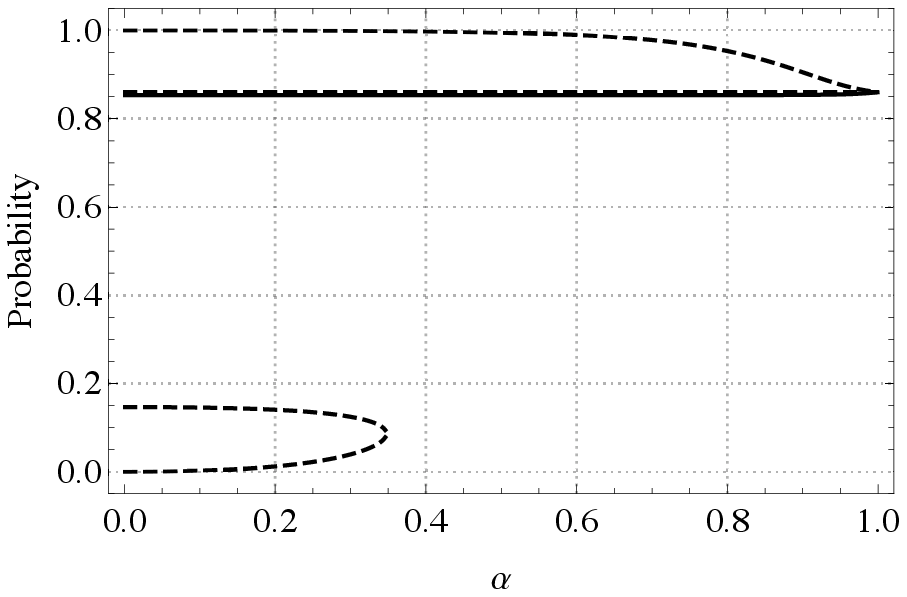}%
            \end{minipage}
            \begin{minipage}{0.31465\textwidth}
                \includegraphics[width=57.81mm,clip,trim=13 13 0 0]{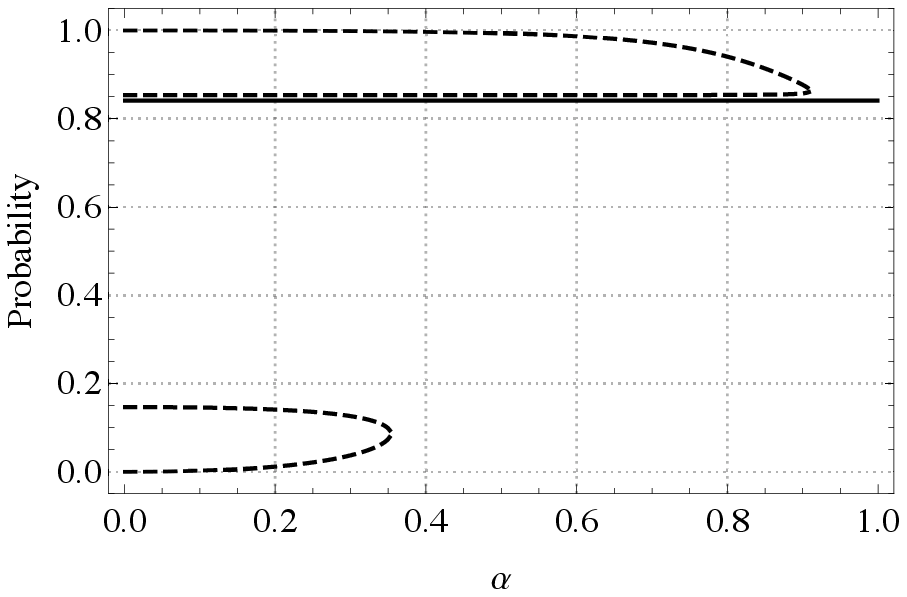}%
            \end{minipage} \\
            \vspace{0.5ex}
            \begin{minipage}{0.33\textwidth}
                \includegraphics[width=60.63mm]{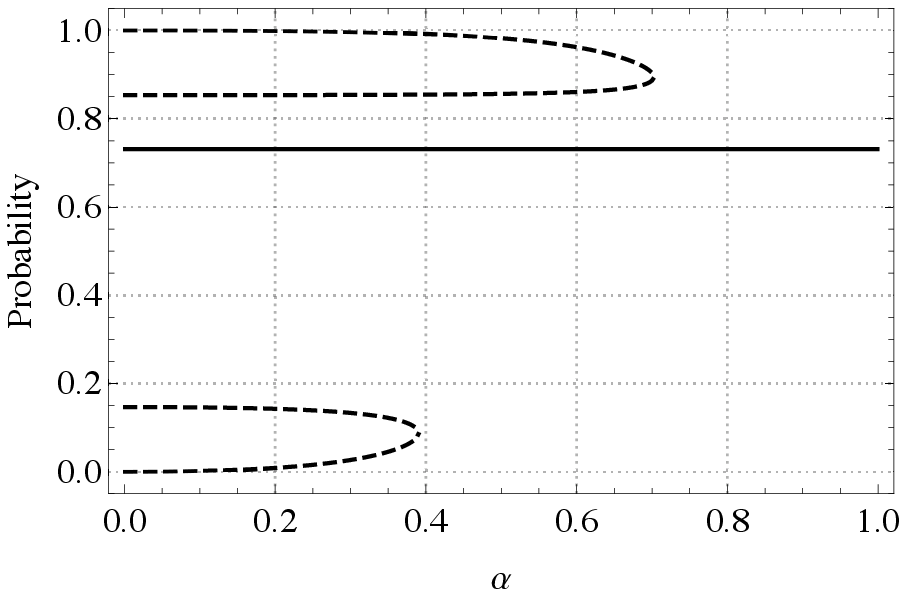}%
            \end{minipage}
            \begin{minipage}{0.31465\textwidth}
                \includegraphics[width=57.81mm,clip,trim=13 0 0 0]{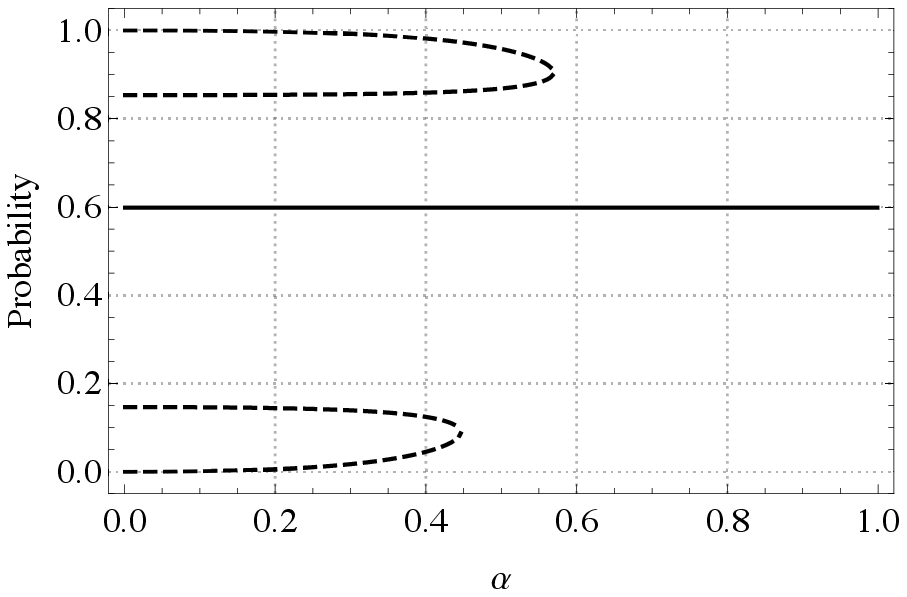}%
            \end{minipage}
            \begin{minipage}{0.31465\textwidth}
                \includegraphics[width=57.81mm,clip,trim=13 0 0 0]{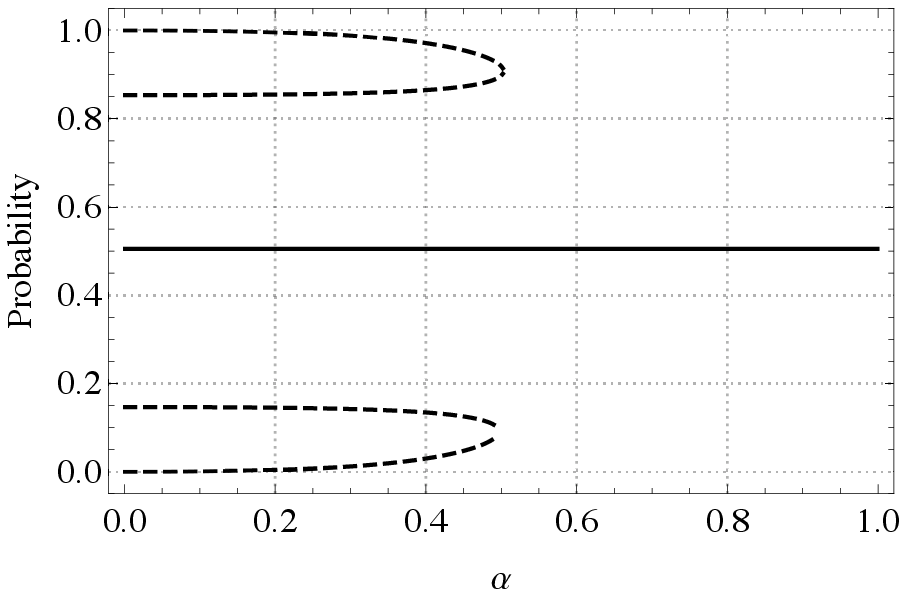}%
            \end{minipage}
        \end{tabular}
        \caption{
            \label{fig_TLS_T02to100}
            Stationary state probability $\rho_{11}$ as a function of $\alpha$ at (going from left to right and top to bottom) $T = 0.2, 0.5, 0.9, 1.0, 1.1, 1.2, 2, 5$, and $100$ with $h = \Gamma = 1$.
            The solid and dashed lines represent the stable and unstable solutions, respectively.
        }
    \end{figure*}
    \begin{figure*}[htb]
        \centering
        \begin{tabular}{l}
            \begin{minipage}{0.33\textwidth}
                \includegraphics[width=60.62mm,clip,trim=0 20 0 0]{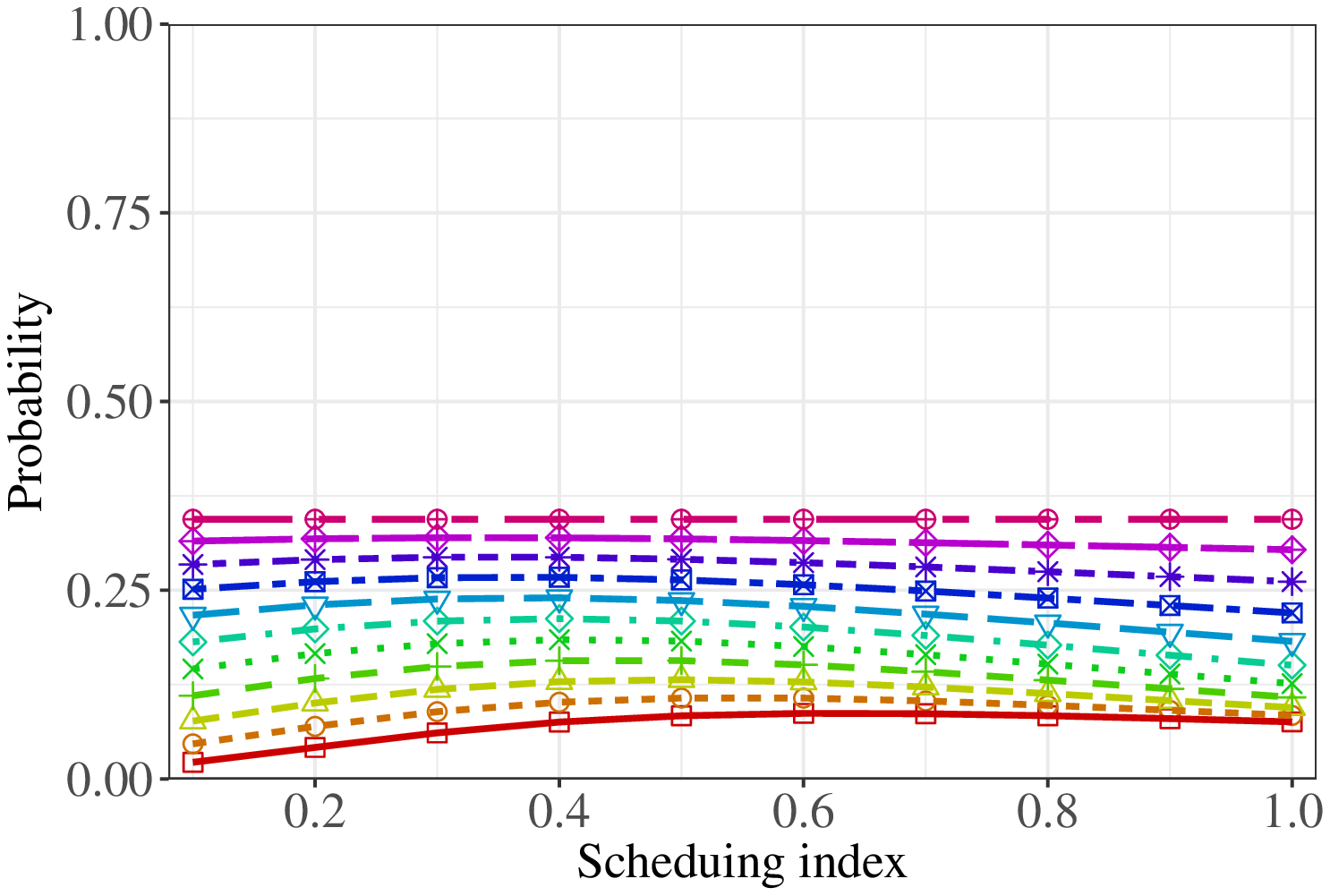}%
            \end{minipage}
            \begin{minipage}{0.31465\textwidth}
                \includegraphics[width=57.81mm,clip,trim=20 20 0 0]{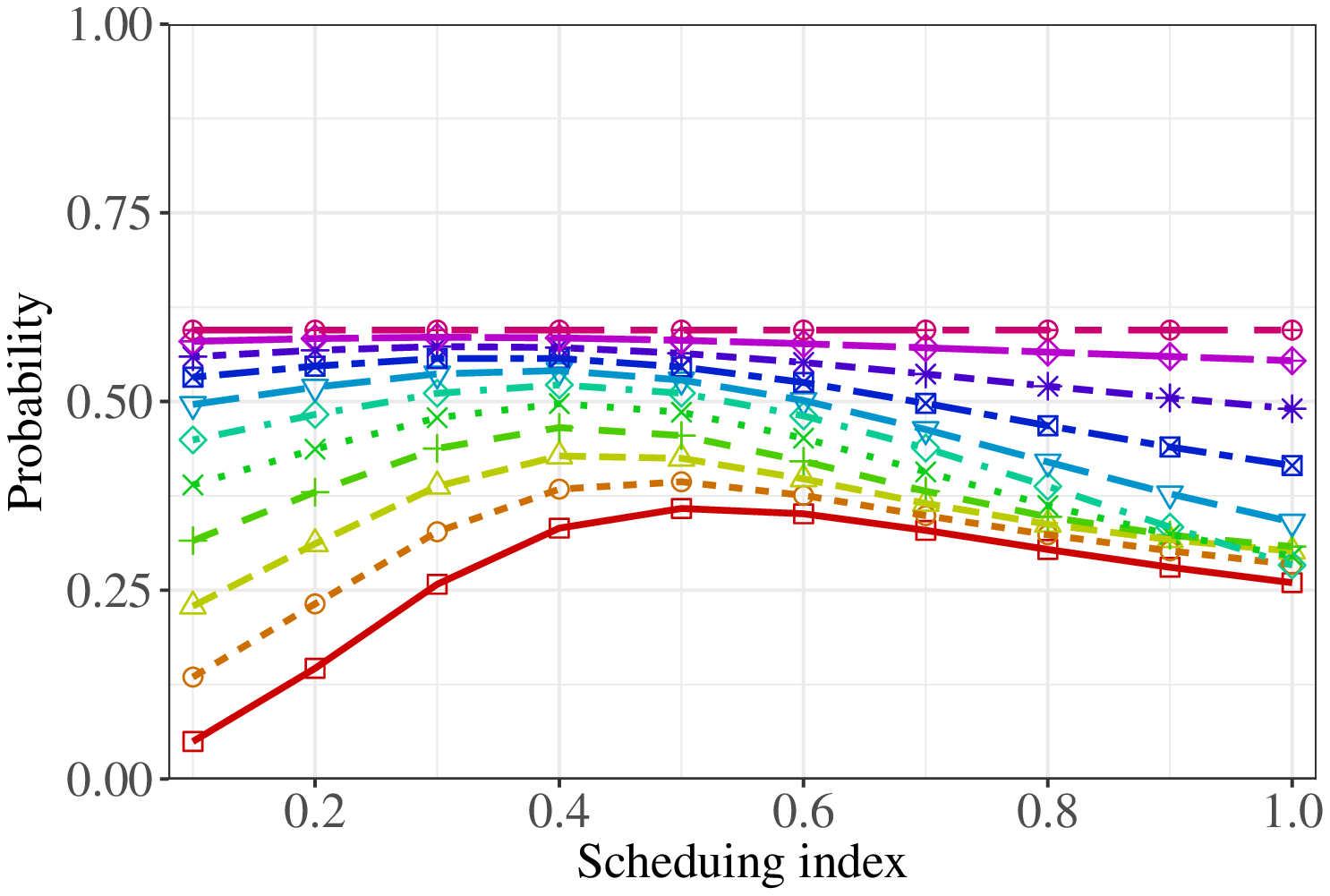}%
            \end{minipage}
            \begin{minipage}{0.31465\textwidth}
                \includegraphics[width=57.81mm,clip,trim=20 20 0 0]{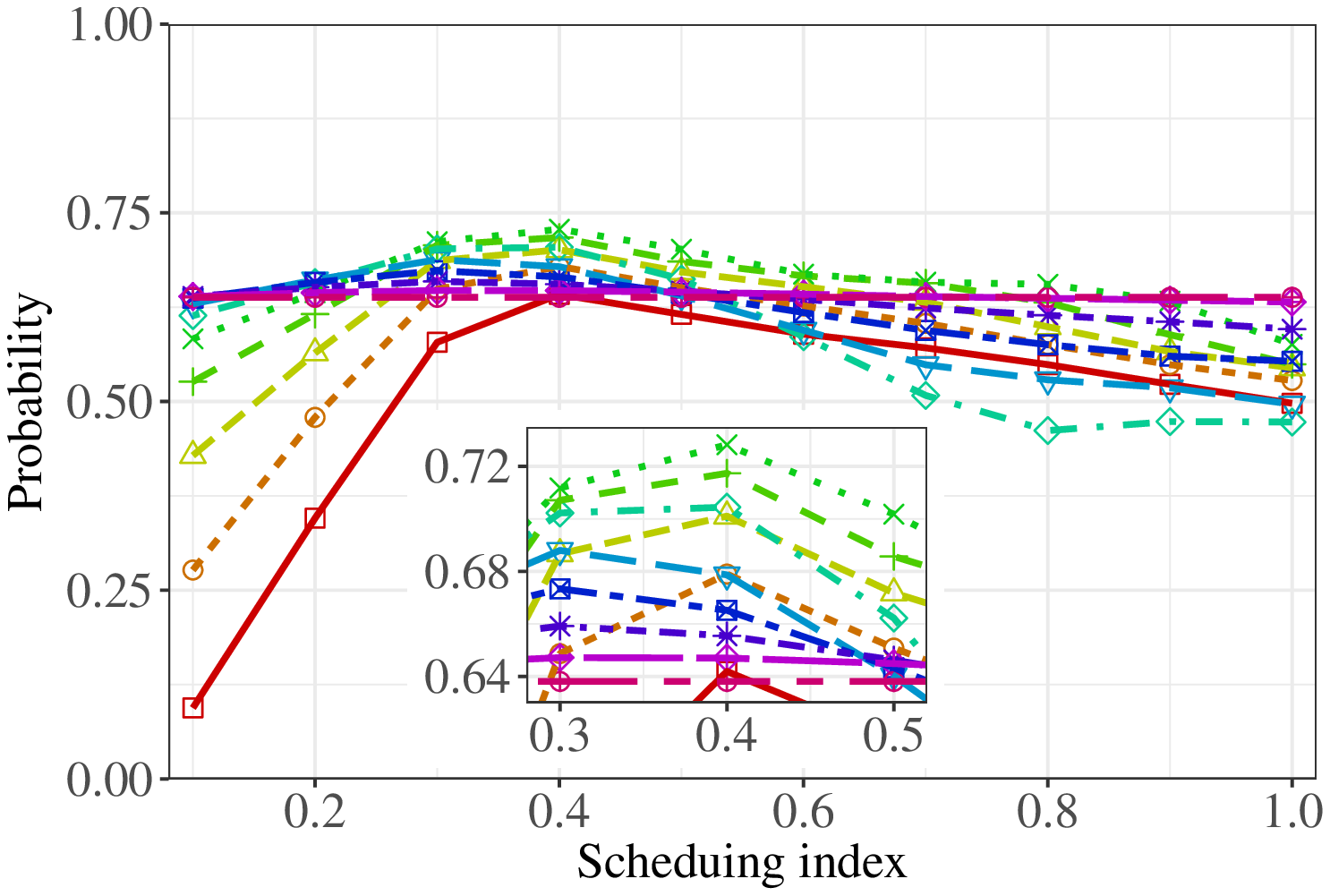}%
            \end{minipage} \\
            \vspace{0.5ex}
            \begin{minipage}{0.33\textwidth}
                \includegraphics[width=60.62mm,clip,trim=0 0 0 0]{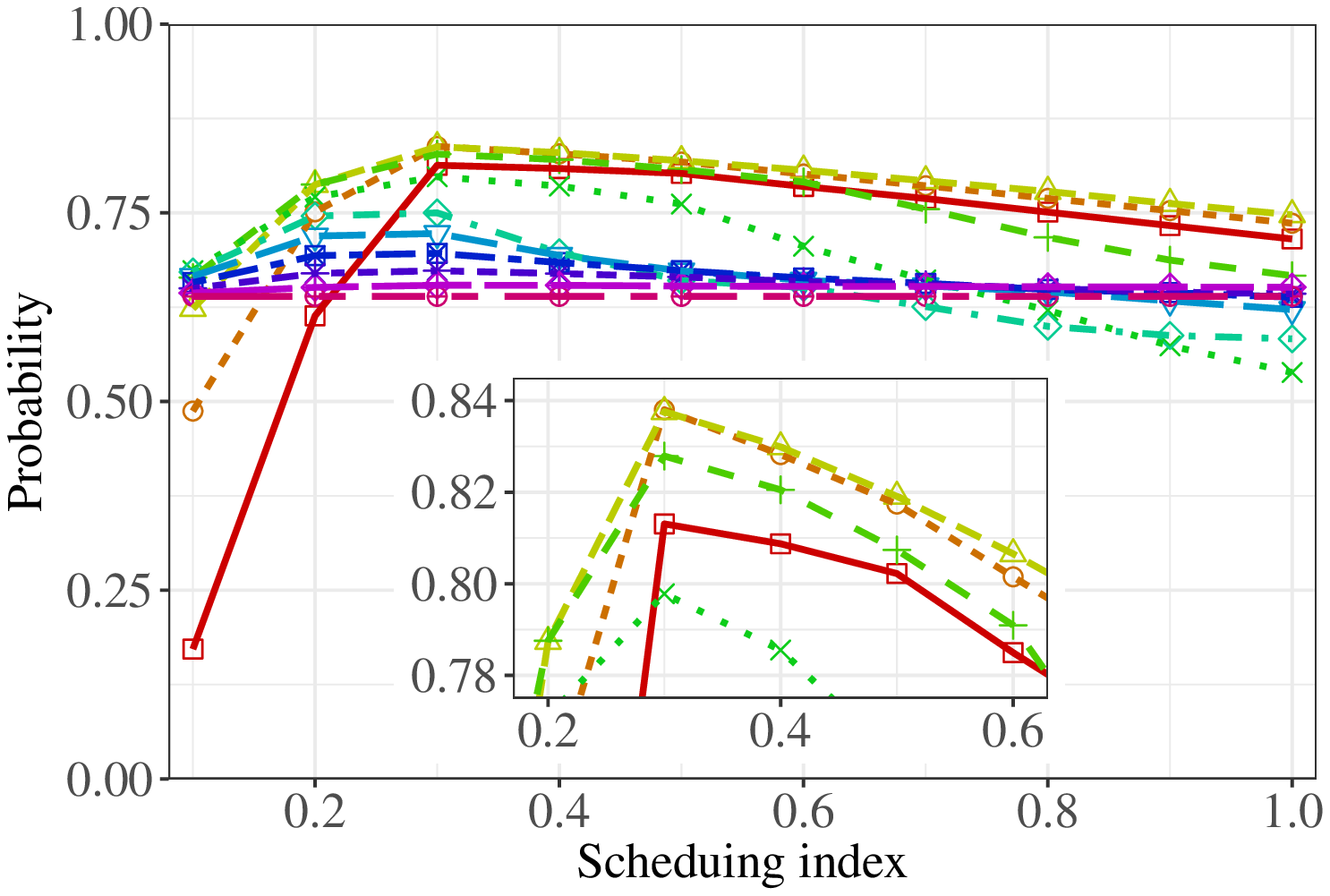}%
            \end{minipage}
            \begin{minipage}{0.31465\textwidth}
                \includegraphics[width=57.81mm,clip,trim=20 0 0 0]{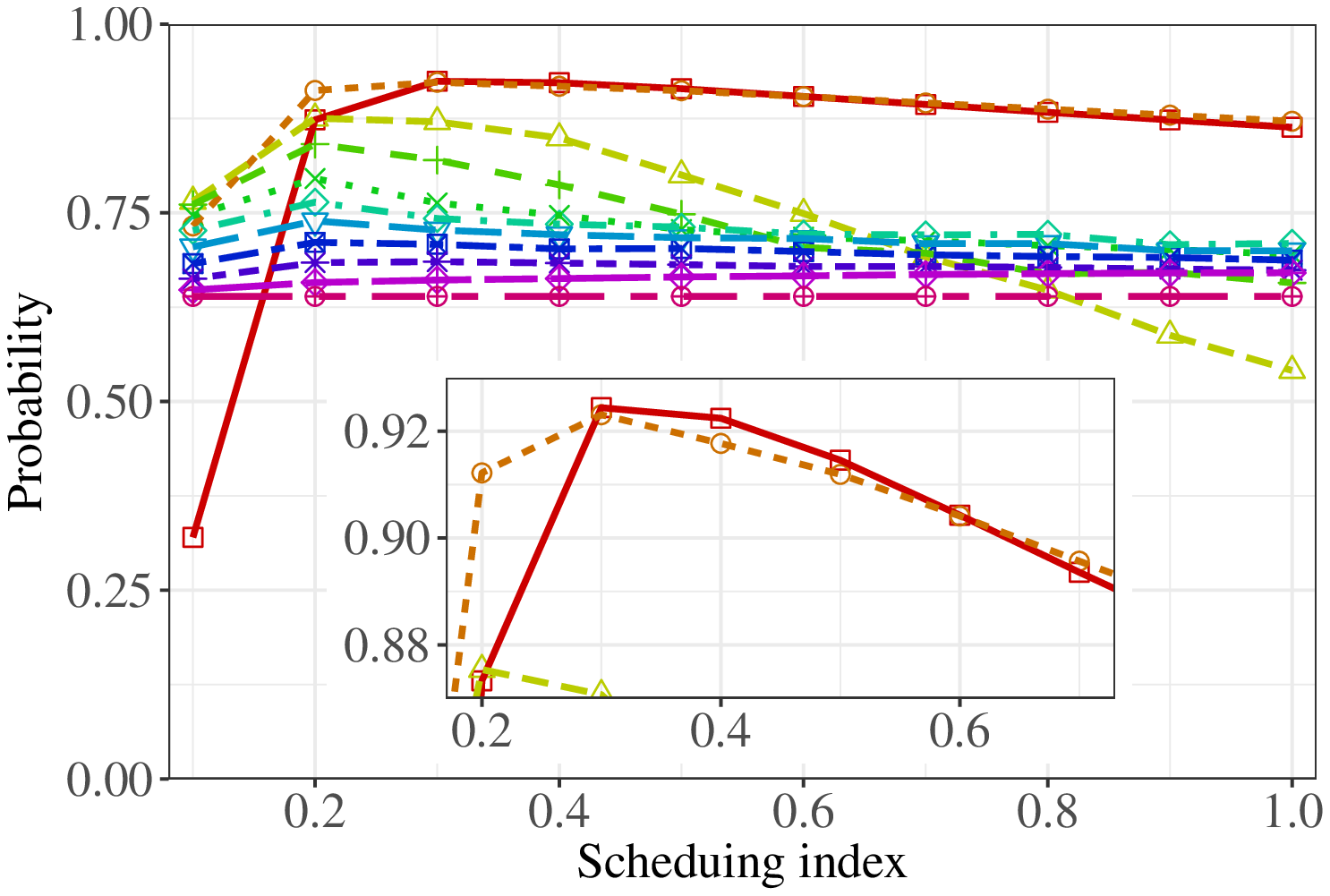}%
            \end{minipage}
            \begin{minipage}{0.31465\textwidth}
                \includegraphics[width=57.81mm,clip,trim=170 80 0 80]{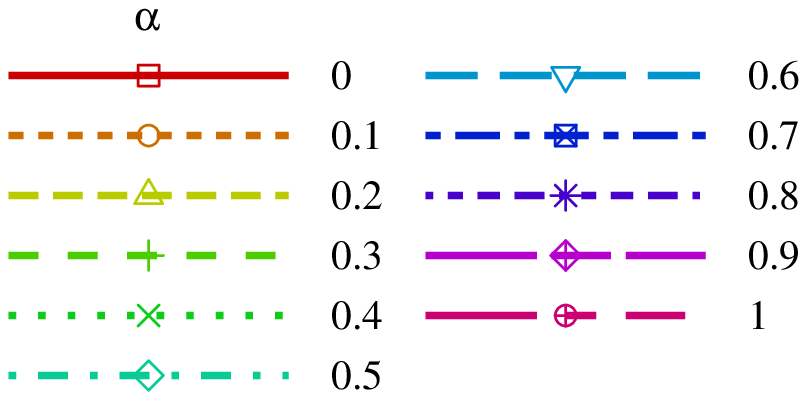}%
            \end{minipage}
        \end{tabular}
        \caption{
            \label{fig_SK_t02to50}
            QUBO ground-state probability as a function of the scheduling index $\gamma$ for total annealing times of (going from left to right and top to bottom) $\tau = 2, 5, 10, 20$, and $50$.
            Each figure depicts the average probabilities for 50 randomly-generated SK model Hamiltonians with mixing parameters $\alpha = 0.1, 0.2, \dots, 1$.
            The inset figures depict magnified views around the tops of the curves.
        }
    \end{figure*}
    \end{widetext}

    \bibliography{ref.bib}

\begin{thebibliography}{35}%
\makeatletter
\providecommand \@ifxundefined [1]{%
 \@ifx{#1\undefined}
}%
\providecommand \@ifnum [1]{%
 \ifnum #1\expandafter \@firstoftwo
 \else \expandafter \@secondoftwo
 \fi
}%
\providecommand \@ifx [1]{%
 \ifx #1\expandafter \@firstoftwo
 \else \expandafter \@secondoftwo
 \fi
}%
\providecommand \natexlab [1]{#1}%
\providecommand \enquote  [1]{``#1''}%
\providecommand \bibnamefont  [1]{#1}%
\providecommand \bibfnamefont [1]{#1}%
\providecommand \citenamefont [1]{#1}%
\providecommand \href@noop [0]{\@secondoftwo}%
\providecommand \href [0]{\begingroup \@sanitize@url \@href}%
\providecommand \@href[1]{\@@startlink{#1}\@@href}%
\providecommand \@@href[1]{\endgroup#1\@@endlink}%
\providecommand \@sanitize@url [0]{\catcode `\\12\catcode `\$12\catcode
  `\&12\catcode `\#12\catcode `\^12\catcode `\_12\catcode `\%12\relax}%
\providecommand \@@startlink[1]{}%
\providecommand \@@endlink[0]{}%
\providecommand \url  [0]{\begingroup\@sanitize@url \@url }%
\providecommand \@url [1]{\endgroup\@href {#1}{\urlprefix }}%
\providecommand \urlprefix  [0]{URL }%
\providecommand \Eprint [0]{\href }%
\providecommand \doibase [0]{http://dx.doi.org/}%
\providecommand \selectlanguage [0]{\@gobble}%
\providecommand \bibinfo  [0]{\@secondoftwo}%
\providecommand \bibfield  [0]{\@secondoftwo}%
\providecommand \translation [1]{[#1]}%
\providecommand \BibitemOpen [0]{}%
\providecommand \bibitemStop [0]{}%
\providecommand \bibitemNoStop [0]{.\EOS\space}%
\providecommand \EOS [0]{\spacefactor3000\relax}%
\providecommand \BibitemShut  [1]{\csname bibitem#1\endcsname}%
\let\auto@bib@innerbib\@empty
\bibitem [{\citenamefont {Kadowaki}\ and\ \citenamefont
  {Nishimori}(1998)}]{Kadowaki1998}%
  \BibitemOpen
  \bibfield  {author} {\bibinfo {author} {\bibfnamefont {Tadashi}\ \bibnamefont
  {Kadowaki}}\ and\ \bibinfo {author} {\bibfnamefont {Hidetoshi}\ \bibnamefont
  {Nishimori}},\ }\bibfield  {title} {\enquote {\bibinfo {title} {{Quantum
  annealing in the transverse Ising model}},}\ }\href {\doibase
  10.1103/PhysRevE.58.5355} {\bibfield  {journal} {\bibinfo  {journal}
  {Physical Review E}\ }\textbf {\bibinfo {volume} {58}},\ \bibinfo {pages}
  {5355--5363} (\bibinfo {year} {1998})},\ \Eprint
  {http://arxiv.org/abs/9804280} {arXiv:9804280 [cond-mat]} \BibitemShut
  {NoStop}%
\bibitem [{\citenamefont {Kadowaki}(1998)}]{Kadowaki1998a}%
  \BibitemOpen
  \bibfield  {author} {\bibinfo {author} {\bibfnamefont {Tadashi}\ \bibnamefont
  {Kadowaki}},\ }\emph {\bibinfo {title} {{Study of Optimization Problems by
  Quantum Annealing}}},\ \href@noop {} {Ph.D. thesis},\ \bibinfo  {school}
  {Tokyo Institute of Technology} (\bibinfo {year} {1998}),\ \Eprint
  {http://arxiv.org/abs/0205020} {arXiv:0205020 [quant-ph]} \BibitemShut
  {NoStop}%
\bibitem [{\citenamefont {Brooke}(1999)}]{Brooke1999}%
  \BibitemOpen
  \bibfield  {author} {\bibinfo {author} {\bibfnamefont {J.}~\bibnamefont
  {Brooke}},\ }\bibfield  {title} {\enquote {\bibinfo {title} {{Quantum
  Annealing of a Disordered Magnet}},}\ }\href {\doibase
  10.1126/science.284.5415.779} {\bibfield  {journal} {\bibinfo  {journal}
  {Science}\ }\textbf {\bibinfo {volume} {284}},\ \bibinfo {pages} {779--781}
  (\bibinfo {year} {1999})},\ \Eprint {http://arxiv.org/abs/0105238}
  {arXiv:0105238 [cond-mat]} \BibitemShut {NoStop}%
\bibitem [{\citenamefont {Farhi}\ \emph {et~al.}(2001)\citenamefont {Farhi},
  \citenamefont {Goldstone}, \citenamefont {Gutmann}, \citenamefont {Lapan},
  \citenamefont {Lundgren},\ and\ \citenamefont {Preda}}]{Farhi2001}%
  \BibitemOpen
  \bibfield  {author} {\bibinfo {author} {\bibfnamefont {Edward}\ \bibnamefont
  {Farhi}}, \bibinfo {author} {\bibfnamefont {Jeffrey}\ \bibnamefont
  {Goldstone}}, \bibinfo {author} {\bibfnamefont {Sam}\ \bibnamefont
  {Gutmann}}, \bibinfo {author} {\bibfnamefont {Joshua}\ \bibnamefont {Lapan}},
  \bibinfo {author} {\bibfnamefont {Andrew}\ \bibnamefont {Lundgren}}, \ and\
  \bibinfo {author} {\bibfnamefont {Daniel}\ \bibnamefont {Preda}},\ }\bibfield
   {title} {\enquote {\bibinfo {title} {{A Quantum Adiabatic Evolution
  Algorithm Applied to Random Instances of an NP-Complete Problem}},}\ }\href
  {\doibase 10.1126/science.1057726} {\bibfield  {journal} {\bibinfo  {journal}
  {Science}\ }\textbf {\bibinfo {volume} {292}},\ \bibinfo {pages} {472--475}
  (\bibinfo {year} {2001})},\ \Eprint {http://arxiv.org/abs/0104129}
  {arXiv:0104129 [quant-ph]} \BibitemShut {NoStop}%
\bibitem [{\citenamefont {Santoro}(2002)}]{Santoro2002}%
  \BibitemOpen
  \bibfield  {author} {\bibinfo {author} {\bibfnamefont {Giuseppe~E.}\
  \bibnamefont {Santoro}},\ }\bibfield  {title} {\enquote {\bibinfo {title}
  {{Theory of Quantum Annealing of an Ising Spin Glass}},}\ }\href {\doibase
  10.1126/science.1068774} {\bibfield  {journal} {\bibinfo  {journal}
  {Science}\ }\textbf {\bibinfo {volume} {295}},\ \bibinfo {pages} {2427--2430}
  (\bibinfo {year} {2002})},\ \Eprint {http://arxiv.org/abs/0205280}
  {arXiv:0205280 [cond-mat]} \BibitemShut {NoStop}%
\bibitem [{\citenamefont {Santoro}\ and\ \citenamefont
  {Tosatti}(2006)}]{Santoro2006}%
  \BibitemOpen
  \bibfield  {author} {\bibinfo {author} {\bibfnamefont {Giuseppe~E}\
  \bibnamefont {Santoro}}\ and\ \bibinfo {author} {\bibfnamefont {Erio}\
  \bibnamefont {Tosatti}},\ }\bibfield  {title} {\enquote {\bibinfo {title}
  {{Optimization using quantum mechanics: quantum annealing through adiabatic
  evolution}},}\ }\href {\doibase 10.1088/0305-4470/39/36/R01} {\bibfield
  {journal} {\bibinfo  {journal} {Journal of Physics A: Mathematical and
  General}\ }\textbf {\bibinfo {volume} {39}},\ \bibinfo {pages} {R393--R431}
  (\bibinfo {year} {2006})}\BibitemShut {NoStop}%
\bibitem [{\citenamefont {Das}\ and\ \citenamefont
  {Chakrabarti}(2008)}]{Das2008}%
  \BibitemOpen
  \bibfield  {author} {\bibinfo {author} {\bibfnamefont {Arnab}\ \bibnamefont
  {Das}}\ and\ \bibinfo {author} {\bibfnamefont {Bikas~K.}\ \bibnamefont
  {Chakrabarti}},\ }\bibfield  {title} {\enquote {\bibinfo {title} {{Colloquium
  : Quantum annealing and analog quantum computation}},}\ }\href {\doibase
  10.1103/RevModPhys.80.1061} {\bibfield  {journal} {\bibinfo  {journal}
  {Reviews of Modern Physics}\ }\textbf {\bibinfo {volume} {80}},\ \bibinfo
  {pages} {1061--1081} (\bibinfo {year} {2008})},\ \Eprint
  {http://arxiv.org/abs/0801.2193} {arXiv:0801.2193} \BibitemShut {NoStop}%
\bibitem [{\citenamefont {Morita}\ and\ \citenamefont
  {Nishimori}(2008)}]{Morita2008}%
  \BibitemOpen
  \bibfield  {author} {\bibinfo {author} {\bibfnamefont {Satoshi}\ \bibnamefont
  {Morita}}\ and\ \bibinfo {author} {\bibfnamefont {Hidetoshi}\ \bibnamefont
  {Nishimori}},\ }\bibfield  {title} {\enquote {\bibinfo {title} {{Mathematical
  foundation of quantum annealing}},}\ }\href {\doibase 10.1063/1.2995837}
  {\bibfield  {journal} {\bibinfo  {journal} {Journal of Mathematical Physics}\
  }\textbf {\bibinfo {volume} {49}},\ \bibinfo {pages} {1--47} (\bibinfo {year}
  {2008})},\ \Eprint {http://arxiv.org/abs/0806.1859} {arXiv:0806.1859}
  \BibitemShut {NoStop}%
\bibitem [{\citenamefont {Tanaka}\ \emph {et~al.}(2017)\citenamefont {Tanaka},
  \citenamefont {Tamura},\ and\ \citenamefont {Chakrabarti}}]{Tanaka2017}%
  \BibitemOpen
  \bibfield  {author} {\bibinfo {author} {\bibfnamefont {Shu}\ \bibnamefont
  {Tanaka}}, \bibinfo {author} {\bibfnamefont {Ryo}\ \bibnamefont {Tamura}}, \
  and\ \bibinfo {author} {\bibfnamefont {B.~K.}\ \bibnamefont {Chakrabarti}},\
  }\href@noop {} {\emph {\bibinfo {title} {{Quantum spin glasses, annealing and
  computation}}}}\ (\bibinfo  {publisher} {Cambridge University Press},\
  \bibinfo {year} {2017})\BibitemShut {NoStop}%
\bibitem [{\citenamefont {Lucas}(2014)}]{Lucas2014}%
  \BibitemOpen
  \bibfield  {author} {\bibinfo {author} {\bibfnamefont {Andrew}\ \bibnamefont
  {Lucas}},\ }\bibfield  {title} {\enquote {\bibinfo {title} {{Ising
  formulations of many NP problems}},}\ }\href {\doibase
  10.3389/fphy.2014.00005} {\bibfield  {journal} {\bibinfo  {journal}
  {Frontiers in Physics}\ }\textbf {\bibinfo {volume} {2}},\ \bibinfo {pages}
  {1--15} (\bibinfo {year} {2014})},\ \Eprint {http://arxiv.org/abs/1302.5843}
  {arXiv:1302.5843} \BibitemShut {NoStop}%
\bibitem [{\citenamefont {Johnson}\ \emph {et~al.}(2011)\citenamefont
  {Johnson}, \citenamefont {Amin}, \citenamefont {Gildert}, \citenamefont
  {Lanting}, \citenamefont {Hamze}, \citenamefont {Dickson}, \citenamefont
  {Harris}, \citenamefont {Berkley}, \citenamefont {Johansson}, \citenamefont
  {Bunyk}, \citenamefont {Chapple}, \citenamefont {Enderud}, \citenamefont
  {Hilton}, \citenamefont {Karimi}, \citenamefont {Ladizinsky}, \citenamefont
  {Ladizinsky}, \citenamefont {Oh}, \citenamefont {Perminov}, \citenamefont
  {Rich}, \citenamefont {Thom}, \citenamefont {Tolkacheva}, \citenamefont
  {Truncik}, \citenamefont {Uchaikin}, \citenamefont {Wang}, \citenamefont
  {Wilson},\ and\ \citenamefont {Rose}}]{Johnson2011}%
  \BibitemOpen
  \bibfield  {author} {\bibinfo {author} {\bibfnamefont {M.~W.}\ \bibnamefont
  {Johnson}}, \bibinfo {author} {\bibfnamefont {M.~H.~S.}\ \bibnamefont
  {Amin}}, \bibinfo {author} {\bibfnamefont {S.}~\bibnamefont {Gildert}},
  \bibinfo {author} {\bibfnamefont {T.}~\bibnamefont {Lanting}}, \bibinfo
  {author} {\bibfnamefont {F.}~\bibnamefont {Hamze}}, \bibinfo {author}
  {\bibfnamefont {N.}~\bibnamefont {Dickson}}, \bibinfo {author} {\bibfnamefont
  {R.}~\bibnamefont {Harris}}, \bibinfo {author} {\bibfnamefont {A.~J.}\
  \bibnamefont {Berkley}}, \bibinfo {author} {\bibfnamefont {J.}~\bibnamefont
  {Johansson}}, \bibinfo {author} {\bibfnamefont {P.}~\bibnamefont {Bunyk}},
  \bibinfo {author} {\bibfnamefont {E.~M.}\ \bibnamefont {Chapple}}, \bibinfo
  {author} {\bibfnamefont {C.}~\bibnamefont {Enderud}}, \bibinfo {author}
  {\bibfnamefont {J.~P.}\ \bibnamefont {Hilton}}, \bibinfo {author}
  {\bibfnamefont {K.}~\bibnamefont {Karimi}}, \bibinfo {author} {\bibfnamefont
  {E.}~\bibnamefont {Ladizinsky}}, \bibinfo {author} {\bibfnamefont
  {N.}~\bibnamefont {Ladizinsky}}, \bibinfo {author} {\bibfnamefont
  {T.}~\bibnamefont {Oh}}, \bibinfo {author} {\bibfnamefont {I.}~\bibnamefont
  {Perminov}}, \bibinfo {author} {\bibfnamefont {C.}~\bibnamefont {Rich}},
  \bibinfo {author} {\bibfnamefont {M.~C.}\ \bibnamefont {Thom}}, \bibinfo
  {author} {\bibfnamefont {E.}~\bibnamefont {Tolkacheva}}, \bibinfo {author}
  {\bibfnamefont {C.~J.~S.}\ \bibnamefont {Truncik}}, \bibinfo {author}
  {\bibfnamefont {S.}~\bibnamefont {Uchaikin}}, \bibinfo {author}
  {\bibfnamefont {J.}~\bibnamefont {Wang}}, \bibinfo {author} {\bibfnamefont
  {B.}~\bibnamefont {Wilson}}, \ and\ \bibinfo {author} {\bibfnamefont
  {G.}~\bibnamefont {Rose}},\ }\bibfield  {title} {\enquote {\bibinfo {title}
  {{Quantum annealing with manufactured spins}},}\ }\href {\doibase
  10.1038/nature10012} {\bibfield  {journal} {\bibinfo  {journal} {Nature}\
  }\textbf {\bibinfo {volume} {473}},\ \bibinfo {pages} {194--198} (\bibinfo
  {year} {2011})},\ \Eprint {http://arxiv.org/abs/1510.06356}
  {arXiv:1510.06356} \BibitemShut {NoStop}%
\bibitem [{\citenamefont {Boixo}\ \emph {et~al.}(2013)\citenamefont {Boixo},
  \citenamefont {Albash}, \citenamefont {Spedalieri}, \citenamefont
  {Chancellor},\ and\ \citenamefont {Lidar}}]{Boixo2013}%
  \BibitemOpen
  \bibfield  {author} {\bibinfo {author} {\bibfnamefont {Sergio}\ \bibnamefont
  {Boixo}}, \bibinfo {author} {\bibfnamefont {Tameem}\ \bibnamefont {Albash}},
  \bibinfo {author} {\bibfnamefont {Federico~M.}\ \bibnamefont {Spedalieri}},
  \bibinfo {author} {\bibfnamefont {Nicholas}\ \bibnamefont {Chancellor}}, \
  and\ \bibinfo {author} {\bibfnamefont {Daniel~A.}\ \bibnamefont {Lidar}},\
  }\bibfield  {title} {\enquote {\bibinfo {title} {{Experimental signature of
  programmable quantum annealing.}}}\ }\href {\doibase 10.1038/ncomms3067}
  {\bibfield  {journal} {\bibinfo  {journal} {Nature communications}\ }\textbf
  {\bibinfo {volume} {4}},\ \bibinfo {pages} {2067} (\bibinfo {year} {2013})},\
  \Eprint {http://arxiv.org/abs/1212.1739} {arXiv:1212.1739} \BibitemShut
  {NoStop}%
\bibitem [{\citenamefont {Albash}\ \emph {et~al.}(2015)\citenamefont {Albash},
  \citenamefont {Vinci}, \citenamefont {Mishra}, \citenamefont {Warburton},\
  and\ \citenamefont {Lidar}}]{Albash2015}%
  \BibitemOpen
  \bibfield  {author} {\bibinfo {author} {\bibfnamefont {Tameem}\ \bibnamefont
  {Albash}}, \bibinfo {author} {\bibfnamefont {Walter}\ \bibnamefont {Vinci}},
  \bibinfo {author} {\bibfnamefont {Anurag}\ \bibnamefont {Mishra}}, \bibinfo
  {author} {\bibfnamefont {Paul~A.}\ \bibnamefont {Warburton}}, \ and\ \bibinfo
  {author} {\bibfnamefont {Daniel~A.}\ \bibnamefont {Lidar}},\ }\bibfield
  {title} {\enquote {\bibinfo {title} {{Consistency tests of classical and
  quantum models for a quantum annealer}},}\ }\href {\doibase
  10.1103/PhysRevA.91.042314} {\bibfield  {journal} {\bibinfo  {journal}
  {Physical Review A}\ }\textbf {\bibinfo {volume} {91}},\ \bibinfo {pages}
  {042314} (\bibinfo {year} {2015})},\ \Eprint {http://arxiv.org/abs/1403.4228}
  {arXiv:1403.4228} \BibitemShut {NoStop}%
\bibitem [{\citenamefont {Amin}\ \emph {et~al.}(2008)\citenamefont {Amin},
  \citenamefont {Love},\ and\ \citenamefont {Truncik}}]{Amin2008}%
  \BibitemOpen
  \bibfield  {author} {\bibinfo {author} {\bibfnamefont {M~H~S}\ \bibnamefont
  {Amin}}, \bibinfo {author} {\bibfnamefont {Peter~J}\ \bibnamefont {Love}}, \
  and\ \bibinfo {author} {\bibfnamefont {C~J~S}\ \bibnamefont {Truncik}},\
  }\bibfield  {title} {\enquote {\bibinfo {title} {{Thermally Assisted
  Adiabatic Quantum Computation}},}\ }\href {\doibase
  10.1103/PhysRevLett.100.060503} {\bibfield  {journal} {\bibinfo  {journal}
  {Physical Review Letters}\ }\textbf {\bibinfo {volume} {100}},\ \bibinfo
  {pages} {060503} (\bibinfo {year} {2008})},\ \Eprint
  {http://arxiv.org/abs/0609332} {arXiv:0609332 [cond-mat]} \BibitemShut
  {NoStop}%
\bibitem [{\citenamefont {Venuti}\ \emph {et~al.}(2017)\citenamefont {Venuti},
  \citenamefont {Albash}, \citenamefont {Marvian}, \citenamefont {Lidar},\ and\
  \citenamefont {Zanardi}}]{Venuti2017}%
  \BibitemOpen
  \bibfield  {author} {\bibinfo {author} {\bibfnamefont {Lorenzo~Campos}\
  \bibnamefont {Venuti}}, \bibinfo {author} {\bibfnamefont {Tameem}\
  \bibnamefont {Albash}}, \bibinfo {author} {\bibfnamefont {Milad}\
  \bibnamefont {Marvian}}, \bibinfo {author} {\bibfnamefont {Daniel}\
  \bibnamefont {Lidar}}, \ and\ \bibinfo {author} {\bibfnamefont {Paolo}\
  \bibnamefont {Zanardi}},\ }\bibfield  {title} {\enquote {\bibinfo {title}
  {{Relaxation versus adiabatic quantum steady-state preparation}},}\ }\href
  {\doibase 10.1103/PhysRevA.95.042302} {\bibfield  {journal} {\bibinfo
  {journal} {Physical Review A}\ }\textbf {\bibinfo {volume} {95}},\ \bibinfo
  {pages} {042302} (\bibinfo {year} {2017})},\ \Eprint
  {http://arxiv.org/abs/1612.07979} {arXiv:1612.07979} \BibitemShut {NoStop}%
\bibitem [{\citenamefont {Dickson}\ \emph {et~al.}(2013)\citenamefont
  {Dickson}, \citenamefont {Johnson}, \citenamefont {Amin}, \citenamefont
  {Harris}, \citenamefont {Altomare}, \citenamefont {Berkley}, \citenamefont
  {Bunyk}, \citenamefont {Cai}, \citenamefont {Chapple}, \citenamefont
  {Chavez}, \citenamefont {Cioata}, \citenamefont {Cirip}, \citenamefont
  {DeBuen}, \citenamefont {Drew-Brook}, \citenamefont {Enderud}, \citenamefont
  {Gildert}, \citenamefont {Hamze}, \citenamefont {Hilton}, \citenamefont
  {Hoskinson}, \citenamefont {Karimi}, \citenamefont {Ladizinsky},
  \citenamefont {Ladizinsky}, \citenamefont {Lanting}, \citenamefont {Mahon},
  \citenamefont {Neufeld}, \citenamefont {Oh}, \citenamefont {Perminov},
  \citenamefont {Petroff}, \citenamefont {Przybysz}, \citenamefont {Rich},
  \citenamefont {Spear}, \citenamefont {Tcaciuc}, \citenamefont {Thom},
  \citenamefont {Tolkacheva}, \citenamefont {Uchaikin}, \citenamefont {Wang},
  \citenamefont {Wilson}, \citenamefont {Merali},\ and\ \citenamefont
  {Rose}}]{Dickson2013}%
  \BibitemOpen
  \bibfield  {author} {\bibinfo {author} {\bibfnamefont {N~G}\ \bibnamefont
  {Dickson}}, \bibinfo {author} {\bibfnamefont {M~W}\ \bibnamefont {Johnson}},
  \bibinfo {author} {\bibfnamefont {M~H}\ \bibnamefont {Amin}}, \bibinfo
  {author} {\bibfnamefont {R}~\bibnamefont {Harris}}, \bibinfo {author}
  {\bibfnamefont {F}~\bibnamefont {Altomare}}, \bibinfo {author} {\bibfnamefont
  {A~J}\ \bibnamefont {Berkley}}, \bibinfo {author} {\bibfnamefont
  {P}~\bibnamefont {Bunyk}}, \bibinfo {author} {\bibfnamefont {J}~\bibnamefont
  {Cai}}, \bibinfo {author} {\bibfnamefont {E~M}\ \bibnamefont {Chapple}},
  \bibinfo {author} {\bibfnamefont {P}~\bibnamefont {Chavez}}, \bibinfo
  {author} {\bibfnamefont {F}~\bibnamefont {Cioata}}, \bibinfo {author}
  {\bibfnamefont {T}~\bibnamefont {Cirip}}, \bibinfo {author} {\bibfnamefont
  {P}~\bibnamefont {DeBuen}}, \bibinfo {author} {\bibfnamefont {M}~\bibnamefont
  {Drew-Brook}}, \bibinfo {author} {\bibfnamefont {C}~\bibnamefont {Enderud}},
  \bibinfo {author} {\bibfnamefont {S}~\bibnamefont {Gildert}}, \bibinfo
  {author} {\bibfnamefont {F}~\bibnamefont {Hamze}}, \bibinfo {author}
  {\bibfnamefont {J~P}\ \bibnamefont {Hilton}}, \bibinfo {author}
  {\bibfnamefont {E}~\bibnamefont {Hoskinson}}, \bibinfo {author}
  {\bibfnamefont {K}~\bibnamefont {Karimi}}, \bibinfo {author} {\bibfnamefont
  {E}~\bibnamefont {Ladizinsky}}, \bibinfo {author} {\bibfnamefont
  {N}~\bibnamefont {Ladizinsky}}, \bibinfo {author} {\bibfnamefont
  {T}~\bibnamefont {Lanting}}, \bibinfo {author} {\bibfnamefont
  {T}~\bibnamefont {Mahon}}, \bibinfo {author} {\bibfnamefont {R}~\bibnamefont
  {Neufeld}}, \bibinfo {author} {\bibfnamefont {T}~\bibnamefont {Oh}}, \bibinfo
  {author} {\bibfnamefont {I}~\bibnamefont {Perminov}}, \bibinfo {author}
  {\bibfnamefont {C}~\bibnamefont {Petroff}}, \bibinfo {author} {\bibfnamefont
  {A}~\bibnamefont {Przybysz}}, \bibinfo {author} {\bibfnamefont
  {C}~\bibnamefont {Rich}}, \bibinfo {author} {\bibfnamefont {P}~\bibnamefont
  {Spear}}, \bibinfo {author} {\bibfnamefont {A}~\bibnamefont {Tcaciuc}},
  \bibinfo {author} {\bibfnamefont {M~C}\ \bibnamefont {Thom}}, \bibinfo
  {author} {\bibfnamefont {E}~\bibnamefont {Tolkacheva}}, \bibinfo {author}
  {\bibfnamefont {S}~\bibnamefont {Uchaikin}}, \bibinfo {author} {\bibfnamefont
  {J}~\bibnamefont {Wang}}, \bibinfo {author} {\bibfnamefont {A~B}\
  \bibnamefont {Wilson}}, \bibinfo {author} {\bibfnamefont {Z}~\bibnamefont
  {Merali}}, \ and\ \bibinfo {author} {\bibfnamefont {G}~\bibnamefont {Rose}},\
  }\bibfield  {title} {\enquote {\bibinfo {title} {{Thermally assisted quantum
  annealing of a 16-qubit problem}},}\ }\href {\doibase 10.1038/ncomms2920}
  {\bibfield  {journal} {\bibinfo  {journal} {Nature Communications}\ }\textbf
  {\bibinfo {volume} {4}},\ \bibinfo {pages} {1903} (\bibinfo {year}
  {2013})}\BibitemShut {NoStop}%
\bibitem [{\citenamefont {Nishimura}\ \emph {et~al.}(2016)\citenamefont
  {Nishimura}, \citenamefont {Nishimori}, \citenamefont {Ochoa},\ and\
  \citenamefont {Katzgraber}}]{Nishimura2016}%
  \BibitemOpen
  \bibfield  {author} {\bibinfo {author} {\bibfnamefont {Kohji}\ \bibnamefont
  {Nishimura}}, \bibinfo {author} {\bibfnamefont {Hidetoshi}\ \bibnamefont
  {Nishimori}}, \bibinfo {author} {\bibfnamefont {Andrew~J.}\ \bibnamefont
  {Ochoa}}, \ and\ \bibinfo {author} {\bibfnamefont {Helmut~G.}\ \bibnamefont
  {Katzgraber}},\ }\bibfield  {title} {\enquote {\bibinfo {title} {{Retrieving
  the ground state of spin glasses using thermal noise: Performance of quantum
  annealing at finite temperatures}},}\ }\href {\doibase
  10.1103/PhysRevE.94.032105} {\bibfield  {journal} {\bibinfo  {journal}
  {Physical Review E}\ }\textbf {\bibinfo {volume} {94}},\ \bibinfo {pages}
  {032105} (\bibinfo {year} {2016})},\ \Eprint
  {http://arxiv.org/abs/1605.03303} {arXiv:1605.03303} \BibitemShut {NoStop}%
\bibitem [{\citenamefont {Albash}\ \emph {et~al.}(2017)\citenamefont {Albash},
  \citenamefont {Martin-Mayor},\ and\ \citenamefont {Hen}}]{Albash2017}%
  \BibitemOpen
  \bibfield  {author} {\bibinfo {author} {\bibfnamefont {Tameem}\ \bibnamefont
  {Albash}}, \bibinfo {author} {\bibfnamefont {Victor}\ \bibnamefont
  {Martin-Mayor}}, \ and\ \bibinfo {author} {\bibfnamefont {Itay}\ \bibnamefont
  {Hen}},\ }\bibfield  {title} {\enquote {\bibinfo {title} {{Temperature
  Scaling Law for Quantum Annealing Optimizers}},}\ }\href {\doibase
  10.1103/PhysRevLett.119.110502} {\bibfield  {journal} {\bibinfo  {journal}
  {Physical Review Letters}\ }\textbf {\bibinfo {volume} {119}},\ \bibinfo
  {pages} {110502} (\bibinfo {year} {2017})},\ \Eprint
  {http://arxiv.org/abs/1703.03871} {arXiv:1703.03871} \BibitemShut {NoStop}%
\bibitem [{\citenamefont {Bloch}(1946)}]{Bloch1946}%
  \BibitemOpen
  \bibfield  {author} {\bibinfo {author} {\bibfnamefont {F.}~\bibnamefont
  {Bloch}},\ }\bibfield  {title} {\enquote {\bibinfo {title} {{Nuclear
  Induction}},}\ }\href {\doibase 10.1103/PhysRev.70.460} {\bibfield  {journal}
  {\bibinfo  {journal} {Physical Review}\ }\textbf {\bibinfo {volume} {70}},\
  \bibinfo {pages} {460--474} (\bibinfo {year} {1946})}\BibitemShut {NoStop}%
\bibitem [{\citenamefont {Arecchi}\ and\ \citenamefont
  {Bonifacio}(1965)}]{Arecchi1965}%
  \BibitemOpen
  \bibfield  {author} {\bibinfo {author} {\bibfnamefont {F.T.}\ \bibnamefont
  {Arecchi}}\ and\ \bibinfo {author} {\bibfnamefont {R.}~\bibnamefont
  {Bonifacio}},\ }\bibfield  {title} {\enquote {\bibinfo {title} {{Theory of
  optical maser amplifiers}},}\ }\href {\doibase 10.1109/JQE.1965.1072212}
  {\bibfield  {journal} {\bibinfo  {journal} {IEEE Journal of Quantum
  Electronics}\ }\textbf {\bibinfo {volume} {1}},\ \bibinfo {pages} {169--178}
  (\bibinfo {year} {1965})}\BibitemShut {NoStop}%
\bibitem [{\citenamefont {Redfield}(1965)}]{Redfield1965}%
  \BibitemOpen
  \bibfield  {author} {\bibinfo {author} {\bibfnamefont {A.G.}\ \bibnamefont
  {Redfield}},\ }\bibfield  {title} {\enquote {\bibinfo {title} {{The Theory of
  Relaxation Processes}},}\ }\href {\doibase
  10.1016/B978-1-4832-3114-3.50007-6} {\bibfield  {journal} {\bibinfo
  {journal} {Advances in Magnetic and Optical Resonance}\ }\textbf {\bibinfo
  {volume} {1}},\ \bibinfo {pages} {1--32} (\bibinfo {year}
  {1965})}\BibitemShut {NoStop}%
\bibitem [{\citenamefont {Kossakowski}(1972)}]{Kossakowski1972}%
  \BibitemOpen
  \bibfield  {author} {\bibinfo {author} {\bibfnamefont {A.}~\bibnamefont
  {Kossakowski}},\ }\bibfield  {title} {\enquote {\bibinfo {title} {{On quantum
  statistical mechanics of non-Hamiltonian systems}},}\ }\href {\doibase
  10.1016/0034-4877(72)90010-9} {\bibfield  {journal} {\bibinfo  {journal}
  {Reports on Mathematical Physics}\ }\textbf {\bibinfo {volume} {3}},\
  \bibinfo {pages} {247--274} (\bibinfo {year} {1972})}\BibitemShut {NoStop}%
\bibitem [{\citenamefont {Metcalf}\ and\ \citenamefont {{Van der
  Straten}}(1999)}]{Metcalf1999}%
  \BibitemOpen
  \bibfield  {author} {\bibinfo {author} {\bibfnamefont {Harold~J.}\
  \bibnamefont {Metcalf}}\ and\ \bibinfo {author} {\bibfnamefont {Peter.}\
  \bibnamefont {{Van der Straten}}},\ }\href@noop {} {\emph {\bibinfo {title}
  {{Laser cooling and trapping}}}}\ (\bibinfo  {publisher} {Springer},\
  \bibinfo {year} {1999})\BibitemShut {NoStop}%
\bibitem [{\citenamefont {Sherrington}\ and\ \citenamefont
  {Kirkpatrick}(1975)}]{Sherrington1975}%
  \BibitemOpen
  \bibfield  {author} {\bibinfo {author} {\bibfnamefont {David}\ \bibnamefont
  {Sherrington}}\ and\ \bibinfo {author} {\bibfnamefont {Scott}\ \bibnamefont
  {Kirkpatrick}},\ }\bibfield  {title} {\enquote {\bibinfo {title} {{Solvable
  Model of a Spin-Glass}},}\ }\href {\doibase 10.1103/PhysRevLett.35.1792}
  {\bibfield  {journal} {\bibinfo  {journal} {Physical Review Letters}\
  }\textbf {\bibinfo {volume} {35}},\ \bibinfo {pages} {1792--1796} (\bibinfo
  {year} {1975})}\BibitemShut {NoStop}%
\bibitem [{\citenamefont {Farhi}\ \emph {et~al.}(2002)\citenamefont {Farhi},
  \citenamefont {Goldstone},\ and\ \citenamefont {Gutmann}}]{Farhi2002}%
  \BibitemOpen
  \bibfield  {author} {\bibinfo {author} {\bibfnamefont {Edward}\ \bibnamefont
  {Farhi}}, \bibinfo {author} {\bibfnamefont {Jeffrey}\ \bibnamefont
  {Goldstone}}, \ and\ \bibinfo {author} {\bibfnamefont {Sam}\ \bibnamefont
  {Gutmann}},\ }\href@noop {} {\emph {\bibinfo {title} {{Quantum Adiabatic
  Evolution Algorithms with Different Paths}}}},\ \bibinfo {type} {Tech. Rep.}\
  (\bibinfo {year} {2002})\ \Eprint {http://arxiv.org/abs/0208135}
  {arXiv:0208135 [quant-ph]} \BibitemShut {NoStop}%
\bibitem [{\citenamefont {Seki}\ and\ \citenamefont
  {Nishimori}(2012)}]{Seki2012}%
  \BibitemOpen
  \bibfield  {author} {\bibinfo {author} {\bibfnamefont {Yuya}\ \bibnamefont
  {Seki}}\ and\ \bibinfo {author} {\bibfnamefont {Hidetoshi}\ \bibnamefont
  {Nishimori}},\ }\bibfield  {title} {\enquote {\bibinfo {title} {{Quantum
  annealing with antiferromagnetic fluctuations}},}\ }\href {\doibase
  10.1103/PhysRevE.85.051112} {\bibfield  {journal} {\bibinfo  {journal}
  {Physical Review E - Statistical, Nonlinear, and Soft Matter Physics}\
  }\textbf {\bibinfo {volume} {85}},\ \bibinfo {pages} {1--8} (\bibinfo {year}
  {2012})},\ \Eprint {http://arxiv.org/abs/1203.2418v2} {arXiv:1203.2418v2}
  \BibitemShut {NoStop}%
\bibitem [{\citenamefont {Seoane}\ and\ \citenamefont
  {Nishimori}(2012)}]{Seoane2012}%
  \BibitemOpen
  \bibfield  {author} {\bibinfo {author} {\bibfnamefont {Beatriz}\ \bibnamefont
  {Seoane}}\ and\ \bibinfo {author} {\bibfnamefont {Hidetoshi}\ \bibnamefont
  {Nishimori}},\ }\bibfield  {title} {\enquote {\bibinfo {title} {{Many-body
  transverse interactions in the quantum annealing of the p -spin
  ferromagnet}},}\ }\href {\doibase 10.1088/1751-8113/45/43/435301} {\bibfield
  {journal} {\bibinfo  {journal} {Journal of Physics A: Mathematical and
  Theoretical}\ }\textbf {\bibinfo {volume} {45}},\ \bibinfo {pages} {435301}
  (\bibinfo {year} {2012})},\ \Eprint {http://arxiv.org/abs/1207.2909}
  {arXiv:1207.2909} \BibitemShut {NoStop}%
\bibitem [{\citenamefont {Crosson}\ \emph {et~al.}(2014)\citenamefont
  {Crosson}, \citenamefont {Farhi}, \citenamefont {Lin}, \citenamefont {Lin},\
  and\ \citenamefont {Shor}}]{Crosson2014}%
  \BibitemOpen
  \bibfield  {author} {\bibinfo {author} {\bibfnamefont {Elizabeth}\
  \bibnamefont {Crosson}}, \bibinfo {author} {\bibfnamefont {Edward}\
  \bibnamefont {Farhi}}, \bibinfo {author} {\bibfnamefont {Cedric Yen-Yu}\
  \bibnamefont {Lin}}, \bibinfo {author} {\bibfnamefont {Han-Hsuan}\
  \bibnamefont {Lin}}, \ and\ \bibinfo {author} {\bibfnamefont {Peter}\
  \bibnamefont {Shor}},\ }\bibfield  {title} {\enquote {\bibinfo {title}
  {{Different Strategies for Optimization Using the Quantum Adiabatic
  Algorithm}},}\ }\href@noop {} {\  (\bibinfo {year} {2014})},\ \Eprint
  {http://arxiv.org/abs/1401.7320} {arXiv:1401.7320} \BibitemShut {NoStop}%
\bibitem [{\citenamefont {Seki}\ and\ \citenamefont
  {Nishimori}(2015)}]{Seki2015}%
  \BibitemOpen
  \bibfield  {author} {\bibinfo {author} {\bibfnamefont {Yuya}\ \bibnamefont
  {Seki}}\ and\ \bibinfo {author} {\bibfnamefont {Hidetoshi}\ \bibnamefont
  {Nishimori}},\ }\bibfield  {title} {\enquote {\bibinfo {title} {{Quantum
  annealing with antiferromagnetic transverse interactions for the Hopfield
  model}},}\ }\href {\doibase 10.1088/1751-8113/48/33/335301} {\bibfield
  {journal} {\bibinfo  {journal} {Journal of Physics A: Mathematical and
  Theoretical}\ }\textbf {\bibinfo {volume} {48}},\ \bibinfo {pages} {335301}
  (\bibinfo {year} {2015})},\ \Eprint {http://arxiv.org/abs/1410.0450}
  {arXiv:1410.0450} \BibitemShut {NoStop}%
\bibitem [{\citenamefont {Hormozi}\ \emph {et~al.}(2017)\citenamefont
  {Hormozi}, \citenamefont {Brown}, \citenamefont {Carleo},\ and\ \citenamefont
  {Troyer}}]{Hormozi2017}%
  \BibitemOpen
  \bibfield  {author} {\bibinfo {author} {\bibfnamefont {Layla}\ \bibnamefont
  {Hormozi}}, \bibinfo {author} {\bibfnamefont {Ethan~W.}\ \bibnamefont
  {Brown}}, \bibinfo {author} {\bibfnamefont {Giuseppe}\ \bibnamefont
  {Carleo}}, \ and\ \bibinfo {author} {\bibfnamefont {Matthias}\ \bibnamefont
  {Troyer}},\ }\bibfield  {title} {\enquote {\bibinfo {title} {{Nonstoquastic
  Hamiltonians and quantum annealing of an Ising spin glass}},}\ }\href
  {\doibase 10.1103/PhysRevB.95.184416} {\bibfield  {journal} {\bibinfo
  {journal} {Physical Review B}\ }\textbf {\bibinfo {volume} {95}},\ \bibinfo
  {pages} {184416} (\bibinfo {year} {2017})},\ \Eprint
  {http://arxiv.org/abs/1609.06558} {arXiv:1609.06558} \BibitemShut {NoStop}%
\bibitem [{\citenamefont {Nishimori}\ and\ \citenamefont
  {Takada}(2017)}]{Nishimori2017}%
  \BibitemOpen
  \bibfield  {author} {\bibinfo {author} {\bibfnamefont {Hidetoshi}\
  \bibnamefont {Nishimori}}\ and\ \bibinfo {author} {\bibfnamefont {Kabuki}\
  \bibnamefont {Takada}},\ }\bibfield  {title} {\enquote {\bibinfo {title}
  {{Exponential Enhancement of the Efficiency of Quantum Annealing by
  Non-Stochastic Hamiltonians}},}\ }\href {\doibase 10.3389/fict.2017.00002}
  {\bibfield  {journal} {\bibinfo  {journal} {Frontiers in ICT}\ }\textbf
  {\bibinfo {volume} {4}},\ \bibinfo {pages} {2} (\bibinfo {year} {2017})},\
  \Eprint {http://arxiv.org/abs/1609.03785} {arXiv:1609.03785} \BibitemShut
  {NoStop}%
\bibitem [{\citenamefont {Shin}\ \emph {et~al.}(2014)\citenamefont {Shin},
  \citenamefont {Smith}, \citenamefont {Smolin},\ and\ \citenamefont
  {Vazirani}}]{Shin2014}%
  \BibitemOpen
  \bibfield  {author} {\bibinfo {author} {\bibfnamefont {Seung~Woo}\
  \bibnamefont {Shin}}, \bibinfo {author} {\bibfnamefont {Graeme}\ \bibnamefont
  {Smith}}, \bibinfo {author} {\bibfnamefont {John~A.}\ \bibnamefont {Smolin}},
  \ and\ \bibinfo {author} {\bibfnamefont {Umesh}\ \bibnamefont {Vazirani}},\
  }\bibfield  {title} {\enquote {\bibinfo {title} {{How ``Quantum'' is the
  D-Wave Machine?}}}\ }\href@noop {} {\  (\bibinfo {year} {2014})},\ \Eprint
  {http://arxiv.org/abs/1401.7087} {arXiv:1401.7087} \BibitemShut {NoStop}%
\bibitem [{\citenamefont {Smolin}\ and\ \citenamefont
  {Smith}(2014)}]{Smolin2014}%
  \BibitemOpen
  \bibfield  {author} {\bibinfo {author} {\bibfnamefont {John~A.}\ \bibnamefont
  {Smolin}}\ and\ \bibinfo {author} {\bibfnamefont {Graeme}\ \bibnamefont
  {Smith}},\ }\bibfield  {title} {\enquote {\bibinfo {title} {{Classical
  signature of quantum annealing}},}\ }\href {\doibase 10.3389/fphy.2014.00052}
  {\bibfield  {journal} {\bibinfo  {journal} {Frontiers in Physics}\ }\textbf
  {\bibinfo {volume} {2}},\ \bibinfo {pages} {52} (\bibinfo {year}
  {2014})}\BibitemShut {NoStop}%
\bibitem [{\citenamefont {Srednicki}(1994)}]{Srednicki1994}%
  \BibitemOpen
  \bibfield  {author} {\bibinfo {author} {\bibfnamefont {Mark}\ \bibnamefont
  {Srednicki}},\ }\bibfield  {title} {\enquote {\bibinfo {title} {{Chaos and
  quantum thermalization}},}\ }\href {\doibase 10.1103/PhysRevE.50.888}
  {\bibfield  {journal} {\bibinfo  {journal} {Physical Review E}\ }\textbf
  {\bibinfo {volume} {50}},\ \bibinfo {pages} {888--901} (\bibinfo {year}
  {1994})},\ \Eprint {http://arxiv.org/abs/9403051} {arXiv:9403051 [cond-mat]}
  \BibitemShut {NoStop}%
\bibitem [{\citenamefont {D'Alessio}\ \emph {et~al.}(2016)\citenamefont
  {D'Alessio}, \citenamefont {Kafri}, \citenamefont {Polkovnikov},\ and\
  \citenamefont {Rigol}}]{DAlessio2016}%
  \BibitemOpen
  \bibfield  {author} {\bibinfo {author} {\bibfnamefont {Luca}\ \bibnamefont
  {D'Alessio}}, \bibinfo {author} {\bibfnamefont {Yariv}\ \bibnamefont
  {Kafri}}, \bibinfo {author} {\bibfnamefont {Anatoli}\ \bibnamefont
  {Polkovnikov}}, \ and\ \bibinfo {author} {\bibfnamefont {Marcos}\
  \bibnamefont {Rigol}},\ }\bibfield  {title} {\enquote {\bibinfo {title}
  {{From quantum chaos and eigenstate thermalization to statistical mechanics
  and thermodynamics}},}\ }\href {\doibase 10.1080/00018732.2016.1198134}
  {\bibfield  {journal} {\bibinfo  {journal} {Advances in Physics}\ }\textbf
  {\bibinfo {volume} {65}},\ \bibinfo {pages} {239--362} (\bibinfo {year}
  {2016})},\ \Eprint {http://arxiv.org/abs/1509.06411} {arXiv:1509.06411}
  \BibitemShut {NoStop}%
\end{thebibliography}%

\end{document}